# Nanoscale Phonon Spectroscopy Reveals Emergent Interface Vibrational Structure of Superlattices

Eric R. Hoglund[1*], De-Liang Bao[2], Andrew O'Hara[2], Sara Makarem[1], Zachary T. Piontkowski[3], Joseph R. Matson[4], Ajay K. Yadav[5], Ryan C. Haisimaier[6], Roman Engel-Herbert[6], Jon F. Ihlefeld[1], Jayakanth Ravichandran[7], Ramamoorthy Ramesh[5], Joshua D. Caldwell[4], Thomas E. Beechem[3,8,9], John A. Tomko[10], Jordan A. Hachtel[11♠], Sokrates T. Pantelides[2,12+], Patrick E. Hopkins[1,10,13°], and James M. Howe[1♦]

[1] Dept. of Materials Science and Engineering, University of Virginia, Charlottesville, VA 22904, USA
[2] Dept. of Physics and Astronomy, Vanderbilt University, Nashville, TN 37235 USA
[3] Sandia National Laboratories, Albuquerque, NM 87123, USA
[4] Dept. of Mechanical Engineering and Electrical Engineering, Vanderbilt University, Nashville TN 37235, USA
[5] Dept. of Materials Science and Engineering, University of California Berkley, Berkley, CA 94720, USA
[6] Dept. of Materials Science and Engineering, Pennsylvania State University, University Park, PA, 16802, USA
[7] Dept. of Chemical Engineering and Materials Science, University of Southern California, Los Angeles, CA 90089, USA
[8] Center for Integrated Nanotechnologies, Sandia National Laboratories, Albuquerque, NM, 87123
[9] School of Mechanical Engineering and the Birck Nanotechnology Center, Purdue University, West Lafayette, Indiana 47907, USA
[10] Dept. of Mechanical and Aerospace Engineering, University of Virginia, Charlottesville, VA 22904, USA
[11] Center for Nanophase Materials Sciences, Oak Ridge National Laboratory, Oak Ridge, TN 37830, USA.
[12] Dept. of Electrical and Computer Engineering, Vanderbilt University, Nashville TN 37235, USA
[13] Dept. of Physics, University of Virginia, Charlottesville, VA 22904, USA

*erh3cq@virginia.edu
♠mailto:hachtelja@ornl.govhachtelja@ornl.gov
+pantelides@vanderbilt.edu
°phopkins@virginia.edu
♦jh9s@virginia.edu

# 1 First paragraph

As the length-scales of materials decrease, heterogeneities associated with interfaces approach the importance of the surrounding materials. This has led to extensive studies of emergent electronic and magnetic interface properties in superlattices.[1–9] However, the interfacial vibrations that impact phonon-mediated properties, like thermal conductivity[10,11], are measured using macroscopic techniques that lack spatial resolution. While it is accepted that intrinsic phonons change near boundaries[12,13], the physical mechanisms and length-scales through which interfacial effects influence materials remain unclear. Herein, we demonstrate the localized vibrational response of interfaces in $SrTiO_3$-$CaTiO_3$ superlattices by combining advanced scanning transmission electron microscopy imaging and spectroscopy, density-functional-theory calculations, and ultrafast optical spectroscopy. Structurally diffuse interfaces are observed that bridge the bounding materials. The local symmetries create phonon modes that determine the global response of the superlattice once the spacing of the interfaces approaches the phonon spatial extent. Our results provide direct visualization of the progression of the local atomic structure and interface vibrations as they come to determine the vibrational response of an entire superlattice. Direct observation of such local atomic and vibrational phenomena demonstrates that their spatial extent needs to be quantified to understand macroscopic behavior. Tailoring interfaces, and knowing their local vibrational response, provides a means of pursuing designer solids having emergent infrared and thermal responses.

# 2 Introduction

The hierarchy of lattices in superlattices presents a tunable phonon-material interaction where, at small to moderate period thicknesses, coherent and localized interface phonons play a major role in controlling properties. The vibrations and coupling regions present at interfaces in superlattices, in a broader context, occur at other interphase and intergranular boundaries and can result in remarkable properties.[3,14–29] Probing vibrations with the lateral spatial resolution required to provide knowledge that can ultimately be used for interface engineering and customization of thermal and infrared properties has remained prohibitively difficult.[10,12,15,30–33] The high spectral and spatial resolution of monochromated electron energy-loss spectroscopy (EELS) in a scanning transmission electron microscope (STEM) provides a unique opportunity to probe the spatial extent of vibrational excitations that are conventionally assessed by infrared light or neutrons. Such resolving capabilities have so far been demonstrated in resolving phonons associated with chemical changes at point defects and stacking faults in crystals.[34–37]

Spatially, modern segmented STEM detectors used for integrated differential phase contrast (iDPC) can image both light and heavy elements, providing knowledge of local symmetry, which dictates vibrational properties.[38–40] Hence, advanced STEM imaging and EELS provides a toolset to understand the intertwined local symmetry and vibrational properties at material interfaces.

In this paper, we combine advanced STEM-iDPC and monochromated EELS experiments with density-functional-theory (DFT) calculations to quantify the local symmetry and vibrational states in $SrTiO_3$-$CaTiO_3$ (STO-CTO) superlattices. We measure the spatial extent of $TiO_6$ octahedral rotation across STO-CTO interfaces (i.e., octahedral coupling) by quantifying the out-of-phase tilt-angle and relate this information to the local Ti and O vibrational response measured with high-spatial-resolution EELS. Second-harmonic-generation measurements (SHG) were performed, which provides a macroscopic measure of both opto-electronic properties and interface density. SHG results agree with our DFT calculations, which are used to model the structural evolution of the superlattices and provide insights into the origins of their differing vibrational states. Finally, through a series of ultrafast optical spectroscopy measurements, we assess the lifetime of zone-center phonon modes in these superlattices, providing insight to the macroscopic property progression observed in these oxide heterostructures. We show that as the superlattice layer thickness approaches the width of structurally diffuse interfaces, where octahedral coupling occurs, the layers lose uniqueness and adopt the structure and vibrational response of the interface. Thus, the vibrational response of the interface becomes more characteristic of the entire material.

## 3  Results

To evaluate the influence of interfaces, we synthesized three STO-CTO superlattices featuring layer thicknesses of 27, 6, 4, 3, and 2 pseudo-cubic unit-cells (SL27, SL6, SL4, SL3, SL2), as shown schematically in Figure 1(a). Large-period SL27 and short-period SL2 were chosen to represent superlattices with well separated and closely spaced interfaces, respectively. SL4 was chosen as an intermediate.

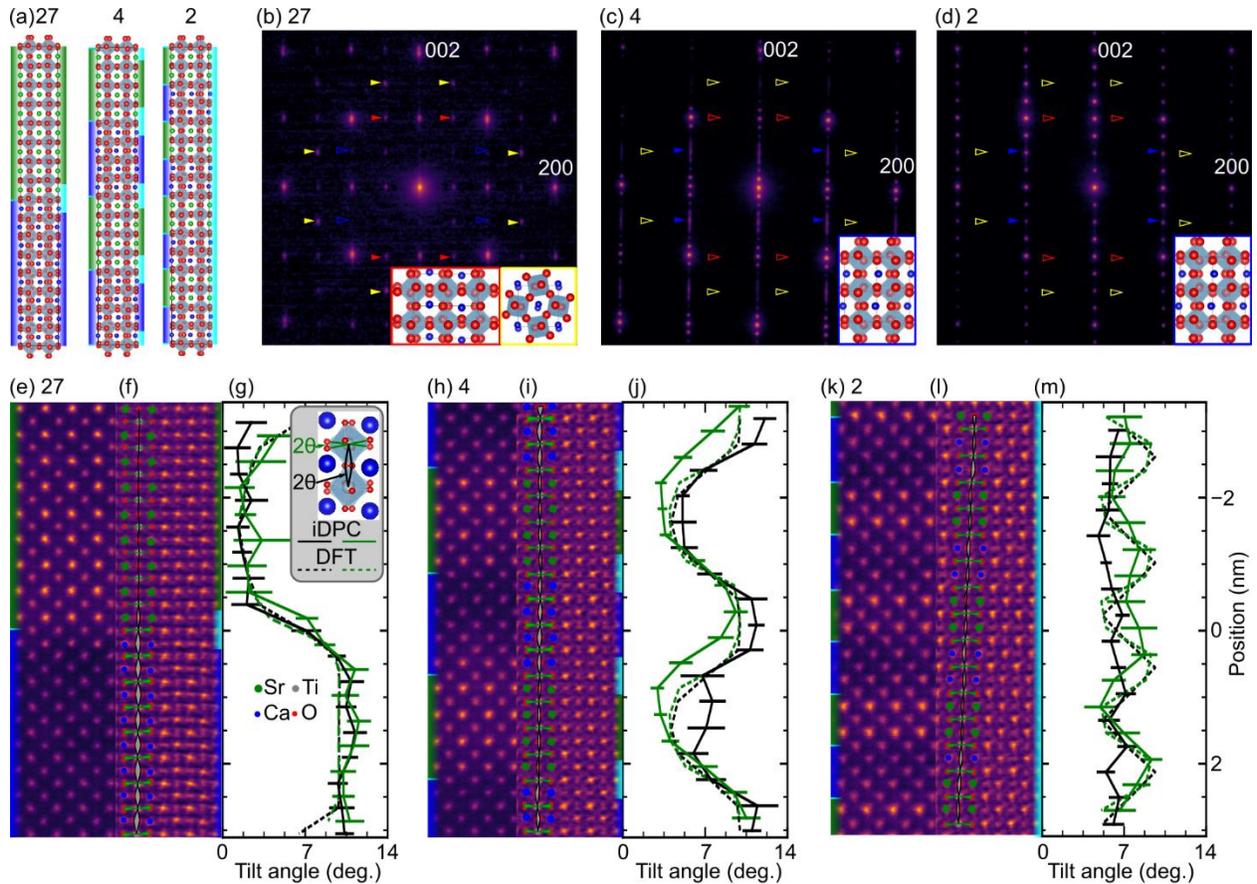

Figure 1. **Period-dependent changes in the symmetry of STO-CTO superlattices.** (a) Superlattice structures calculated from DFT with colored-bar schematics denoting the (left) chemically and (right) structurally defined interfaces. Here green, blue, and cyan rectangles correspond to STO, CTO, and interface layers, respectively; the same colors are used in e,f,h,i,k, and l panels. Green, blue, grey, and red circles in a,e,f,h,i, and k correspond to Sr, Ca, Ti, and O atoms, respectively. [100] zone-axis SADP in (b) SL27, (c) SL4, and (d) SL2 grown on $NdGaO_3$. Colored arrows correspond to ordered reflections from the three possible domains. Solid arrows indicate ordered reflections that exist, and hollow arrows indicate absences. Insets b-d show ball-and-stick models of the orientations present with border colors matching the arrows. Red and blue arrows and insets are viewed along an out-of-phase tilt-axis and the yellow are viewed along an in-phase tilt-axis. In (c) and (d) superlattice reflections are seen in the 001 direction. In (b), closely spaced superlattice reflections appear as streaking of the fundamental reflections. (e,h,k) ADF, (f,i,l) iDPC images, and (g,j,m) octahedral tilt-angles of (e-g) SL27, (h-j) SL4, and (k-m) SL2. The legend inside (g) illustrates the (green) in-plane and (black) out-of-plane tilt-angles. The tilt-angles for a one unit-cell column are overlayed in each iDPC image to demonstrate the changing in-plane (green triangles) and out-of-plane (grey triangles) tilt-angles.

In (g,j,m), solid and dashed curves are from experimental measurements and calculations, respectively. Error bars represent one standard deviation. Chemically abrupt interfaces are illustrated to the left of ADF images (e,h,k) and model structures (a), illustrating the abrupt change between STO (green) and CTO (blue) layers. Chemically diffuse interfaces are illustrated to the right of iDPC images (f,i,l) and model structures (a), illustrating the non-abrupt symmetry changes that are occurring as a result of octahedral coupling.

To quantify the structure of each superlattice, we acquired selected-area electron diffraction patterns (SADP). The SADP reveals that the orientation of octahedral tilts are different in the SL27 structure than in the SL4 and SL2 structures. The large-period SL27 structure exhibits ordered reflections from two of three possible *Pbnm*-CTO domains (Figure 1(b)), each having an in-plane *c*-axis as illustrated by the ball-and-stick insets. Ordered reflections observed from the SL4 and SL2 samples, indicate a single out-of-plane *c*-axis (Figure 1(c,d)). This microstructural transition is accompanied by a relaxation of the lattice parameters to a single intermediate value (Figure S3). Thus, as the layer thickness decreases, the underlying crystal structure adapts, which could be enabled by octahedral tilt.[3,4,19,21–24]

To investigate the potential of such octahedral tilting in the superlattices further, we use annular dark-field (ADF) and iDPC to quantify octahedral tilting (Figure 1(e-m)). The Z-contrast of the ADF images (Figure 1(e,h,k)) allows for discrimination between the brighter Sr and darker Ca atoms and shows chemically abrupt transitions between the two. The iDPC images (Figure 1(f,i,l)) allows the measurement of the positions of oxygen and titanium columns, enabling quantification of the octahedral tilt angles. In-plane (green triangles) and out-of-plane (grey triangles) are overlaid on one unit-cell of each iDPC image to demonstrate the changing octahedral tilt-angles from layer-to-layer and structure-to-structure. For example, the iDPC image of SL27 (Figure 1(f)) has three regions defined by observing the splitting of oxygen columns labeled with red circles; 1) single columns in STO, 2) split columns in CTO, and 3) an intermediate splitting at the interface. The tilts are consistent with simple cubic $Pm\bar{3}m$-STO containing no tilt, orthorhombic $Pbnm$-CTO viewed along an out-of-phase tilt axis, and a region where tilts transition from finite angles in CTO to none in STO. Scanning convergent-beam electron diffraction corroborates these observations (see Figure S4).

To quantify the observed changes in crystal structure with reduced interface separation, the in-plane and out-of-plane octahedral angles are measured (according to the legend in Figure 1(g)). The plane averaged tilt-angle is shown in Figures 1(g,j,m), see supplementary information for quantitative values. A coupling region is present at the STO-CTO interface of SL27 (Figure 1(g)). Similar coupling regions observed in other perovskite heterostructures result in extraordinary electrical and magnetic properties.[3,7–9,19,19–24,41] We loosely define the structurally diffuse interface width as one unit-cell centered at the chemically abrupt $TiO_2$ interface as schematically shown in Figure 1(a).

We then turn to SL4 and SL2 to understand how the octahedra couple across the interfaces when the interface spacing is comparable to the diffuse interface width. The CTO oxygen column splitting in SL4 is less pronounced than in SL27. Some oxygen columns within the STO layers are distinctly split while most appear elliptical from the partial overlap of the splitting column. The octahedra in STO and CTO are entirely coupled, as seen in the sinusoidal profile of the tilt-angles (Figure 1(j)). Coupling is even more apparent in SL2 (Figure 1(m)), where a nearly constant tilt-angle extends throughout the entire structure. Similar titling of STO octahedra has also been observed in short-period $BaTiO_3$-STO superlattices.[15] Here we show that incorporation of the atomic displacements in STO is an interface mediated process.

Using the STEM results we can define three types of superlattices: (1) Long-period superlattices, such as SL27, exhibit monolithic phases with structurally diffuse interfaces. (2) Moderate-periods (SL4) with modified monolithic phases and structurally diffuse interfaces taking up a sizable fraction of the superlattice. (3) Short-periods (SL2) comprise entirely interface regions and are better described as an ordered structure, having a global symmetry characteristic of that seen at the interfaces of all superlattices. In SL2, a tilt-angle near 7°, which exists only at the interface of the large-period superlattice, is now present throughout the entire superlattice and the structural distinction of each layer has vanished. Therefore, the material is more accurately described as having chemically ordered unit-cells with a single tilt-angle and $Sr_2Ca_2Ti_4O_{12}$ basis.[5,42,43,43–45] More simply, the short-period superlattice has become a *crystal of interfaces*.

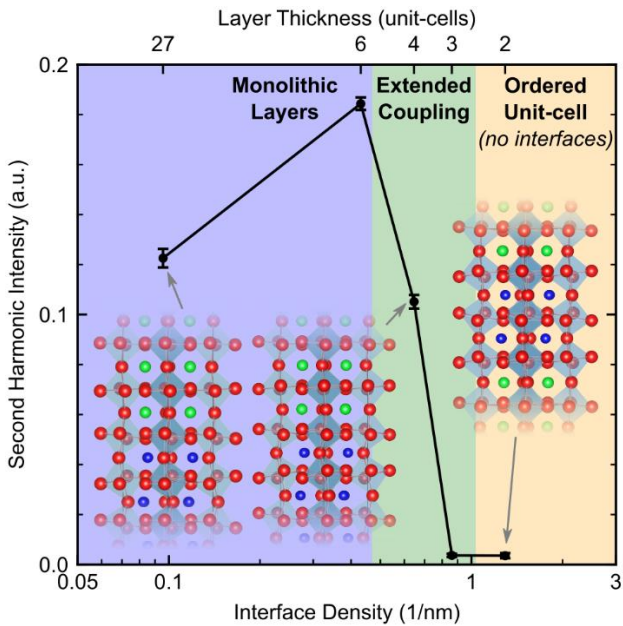

Figure 2: **Second-harmonic intensity indicates short-period superlattices lack interfaces.** Second-harmonic intensity of STO-CTO superlattices with varying periodicity, demonstrating various regimes of structural transitions and their role on electronic/optical properties of heterostructures. The error bars are calculated from the mean square deviation of a parabolic fit to the measured second-harmonic intensity vs incident electric field. Ball-and-stick models are included to pictorially show the connection between octahedral tilt and the presence structurally diffuse interfaces, or lack thereof.

We perform SHG to provide further evidence of the structural regimes, which also lends insight to the electronic properties of the superlattices, as shown in Figure 2. In contrast to linear optical measurements, which are dictated by the average of the linear response of the materials comprising the superlattice[46], the higher rank dielectric tensor associated with SHG vanishes if the constituent materials have inversion symmetry. Both STO and CTO have inversion symmetry, so the SHG intensity of STO-CTO superlattices is directly related to the polarizability of the *interface* where inversion symmetry is broken. An increase in SHG intensity is observed from SL27 to SL6 as the density of interfaces increases and layers remain independent with respect to each other. However, a marked decrease in the nonlinear optical response is observed in SL4 and rapidly vanishes in superlattices with short periodicities; as the heterostructure transitions from independent monolithic layers, to coupled layers, to a single centrosymmetric structure, the second-order optical response approaches zero. In other words, the structural transitions are directly reflected in the polarizability of the superlattices.

To support the observed crystallographic structure of the superlattices and predict the effect of the observed local symmetries on vibrations, we performed DFT calculations on several prototypical superlattice models. Octahedral tilt-angles for both the experimental and theoretical results are in good agreement in both amplitude and periodicity, as indicated by the dashed curves shown in Figure 1(g, j, m). It is noteworthy that the calculations show distinguishable oscillations between octahedral tilts in the STO and CTO layers in SL2, whereas the experimental angles appear decoupled from the chemical identity of the planes. This may be caused by finite amounts of intermixing across the interface (Figure S5(a-e)), which is difficult to prevent experimentally, or a metastable phase (Figure S5(g)). From both the structural calculations and experimental measurements, we can conclude that, as the period thickness decreases, the system converges toward a single, emergent structure.

Since phonon frequencies are affected by changes of bond lengths and bond angles, we expect the evolution of the octahedral tilts in our superlattices (Figure 1) to affect the phonon density of states (DOS), which drives the inherent thermal and infrared optical properties. To evaluate this possibility, we employed DFT to calculate the phonon DOS projected on the oxygen and Ti atoms for each of the three superlattices (Figure 3(a)) (see methods for further description). The PDOS show three peaks: (1) ~37 meV, (2) ~60 meV, and (3) ~97 meV, which are further discussed in SI. From SL27 to SL4 and SL2, (1) and (2) red-shift while (3) blue-shifts, indicating an evolution of the superlattice phonon modes as the layer thickness decreases. Thus, vibrational modes assigned to the octahedra change energy as the octahedra they derive from change tilt-angle with decreasing superlattice period thickness.

The attribution of peak shifts to changes in octahedral tilts is further supported by comparing the PDOS for each constituent layer and for the interface in the three superlattices (Figure 3(b)). We see that in the SL27 the total DOS for the system (black) deviates from the STO (green), CTO (purple) and interface (orange) spectra. However, in SL4 and SL2 the total DOS tracks the interface spectrum almost perfectly, illustrating how the interface dominates these shorter-period structures. In other words, as the layer thickness decreases, both the structural and vibrational state converge toward the respective state of the interfaces.

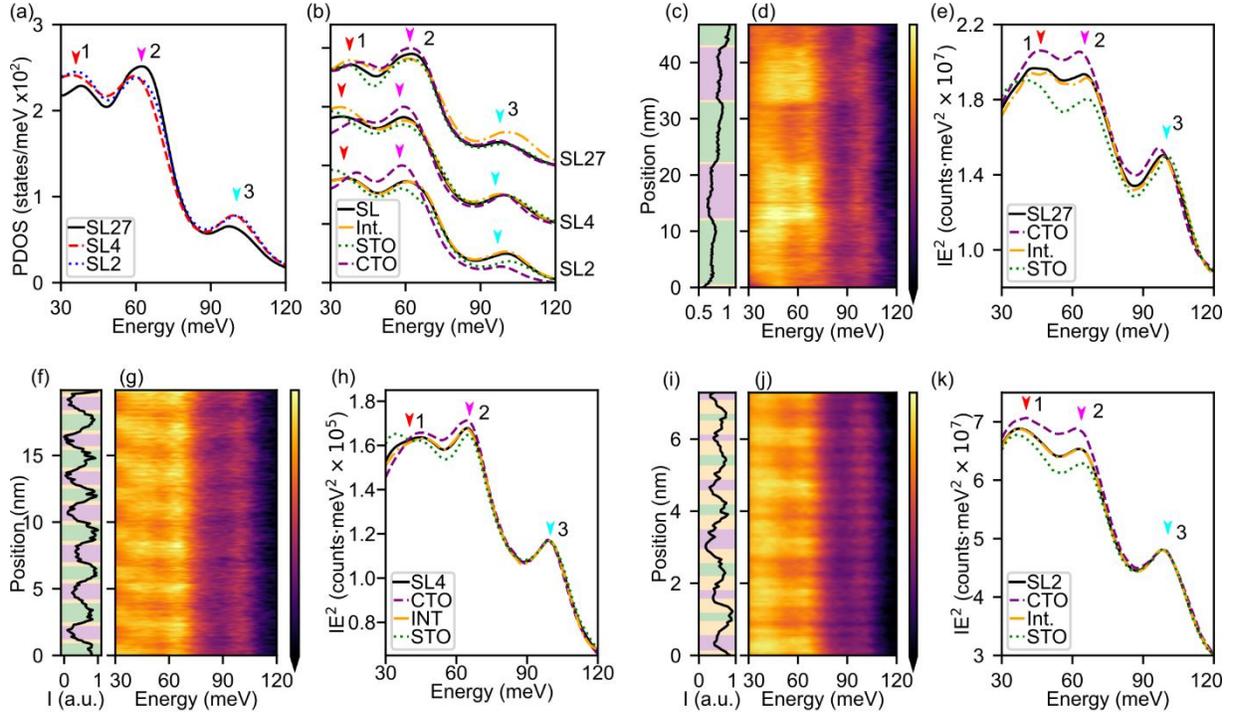

Figure 3. **Localized vibrational response of superlattices indicates emergent role of the interfacial symmetry.** (a) DFT-calculated phonon DOS projected on the octahedron oxygen and Ti atoms of SL27, SL4, and SL2 models. Arrows indicate the dominant phonon peaks. Cascade of (b) DFT-calculated phonon DOS projected on STO (green), CTO (purple), and interface (orange) layers and the total DOS (black) each for superlattice model. (c-k) Monochromated STEM-EELS line profile analyses of the three SL structures: (c-e) SL27, (f-h) SL4, and (i-k) SL2 each with the (c,f,i) ADF profile, (d,g,j) EELS profile, and (e,h,k) integrated spectra from each layer (as indicated by colored regions in the ADF profile).

The evolution in vibrational response of the superlattice is observed experimentally via spatially resolved off-axis vibrational EELS.[35,37] With this approach, the difference in vibrational response within the diffuse interface region can be directly compared to that within the constituent layers themselves. The ADF line profiles of the SL27, SL4, and SL2 are shown in Figure 3(c,f,i), and exhibit a clear distinction between the heavy STO and light CTO layers. The simultaneously acquired EELS are shown in Figure 3(d,g,j). Changes in layer-to-layer response are clearly observed in the superlattices. For example, in SL27 peak 1 is at a lower energy in the STO compared to the CTO, but peaks 2 and 3 are at higher energies, with the interfaces exhibiting intermediate values, demonstrating the capacity to measure changes induced by local atomic displacements. Furthermore, we note that, while the interface and total-structure spectra bear some similarities, there are features in the interface spectra that cannot be reproduced by a mixture of the bulk-like CTO and STO phases (shown in more detail in Figure S11(a,b)). Unique vibrations emerging from the octahedral coupling at the interface must be present to account for the discrepancy between the total and interface spectra, as calculated via DFT. Thus, we demonstrate that variations in the localized vibrational spectra are ascribed to the regions of differing symmetry, namely the STO, CTO, and structurally diffuse interfaces.

Like the octahedral-tilt variation, spatial variations in the EELS response reduce with decreasing period. The spectral similarity from layer-to-layer in SL4 relative to SL27 and the exact match between the total and interface spectra indicate that the vibrational state of the superlattice is approaching that of the interfaces, demonstrating the importance of local vibrational structure as length-scales decrease. The observed response of the superlattice is akin to a crystalline hybrid, where layers hybridize and couple forming new phonons with decreasing layer thickness.[12,47] The global response of the interface vibrations further demonstrates the importance of local vibrational structure as length-scales decrease.

The predominance of these interface vibrations and their global response has been shown to affect the thermal characteristics of STO-CTO superlattices, where a crossover from incoherent to coherent phonon transport is observed as the heterostructure periodicity decreases.[10] Previous reports have suggested that reduced zone-folding of phonon dispersion leads to an increased group velocity, but direct evidence connecting underlying phononic processes, new modes, and structure to the macroscopic transport mechanisms remains lacking.[10,48]

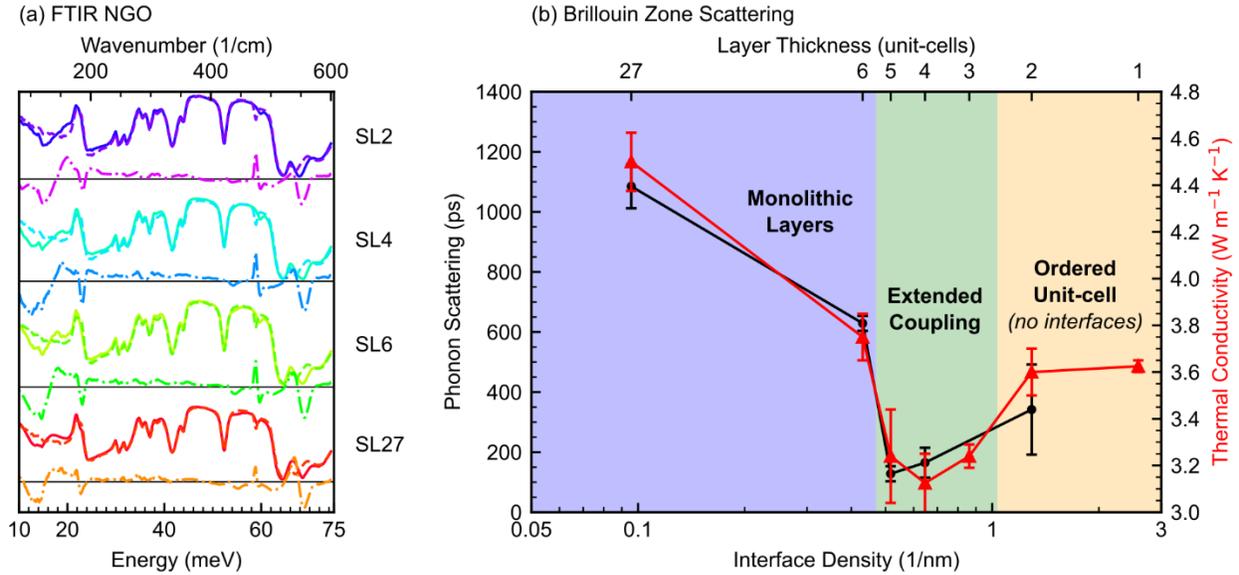

Figure 4: **FTIR and time-domain Brillouin scattering response of STO-CTO superlattices.** (a) Raw (solid), fitted (dashed), and residual (dot-dashed) data for UV-Raman from superlattices on an NGO substrate. 200 nm STO and CTO thin-films on substrates used to fit superlattice spectra are shown as the bottom of the panel. Difference curves are scaled by a factor of two for clarity. (b) Phonon lifetimes (black squares) as measured via time-domain Brillouin scattering in comparison to the thermal conductivity (red triangles, from reference 10) of STO-CTO superlattices with varying periodicities. The strong correlation between the two techniques conclusively demonstrates a transition in phonon scattering rates across the structural transitions elucidated upon with STEM/EELS. The error bars represent the standard deviation.

To connect the localized interface modes and structural transitions observed above to a macroscopic response, we perform Fourier Transform-Infrared spectroscopy (FTIR) (Figure 4(a)).[10,15–18] The residuals from a fitted linear combination of STO and CTO films are quantified to accentuate changes that are unique to the superlattice. Due to selection rules, the spectra of STO contains less reflectivity minima relative to CTO. As the STO and interface incorporate CTO tilt patterns further loss of reflectivity should occur.[49–52] Residual responses are observed near 500 cm$^{-1}$ and 560 cm$^{-1}$ that decrease with decreasing period thickness (Figure 4(a)). UV-Raman experiments show similar trends and are discussed in greater depth in the SI section S5. DFT informs that IR-active phonons emerge in the superlattices and are associated with layer-localized, layer-delocalized, and interface-localized Ti-O vibrational modes that are not IR-active in bulk STO or CTO, as shown in Figure S8. The residual responses are similar in energy to those observed locally with both EELS and DFT. Additionally, the sum of residuals (Figure S16) scales with interface density. We therefore conclude they are a consequence of the local displacements existing at the interfaces of the superlattice.

To directly investigate the emergent phonon dynamics, we perform time-domain Brillouin scattering (TDBS) measurements, which detect the propagation of zone-center longitudinal modes as a function of time, thus providing a measure of their lifetime.[53,54] For short-periods (SL2), an increase in both the phonon scattering time and thermal conductivity is observed; this is no longer described by a superlattice with interfaces, but rather has a single structure with uniform octahedral tilts. We find the thermal conductivity trends reported previously to be strongly correlated to the phonon scattering times (Figure 4), with both decreasing with increasing interface density to a minima at SL4.[10] Thus, in strong agreement with both our high-resolution TEM, EELS, and SHG results, the phonon lifetime is also found to increase as material interfaces vanish. This combination of the structural, electronic/optical, and vibrational characterization techniques unambiguously demonstrates the underlying coupling of heterostructures and emergent global properties that are driven by interfaces.

## 4  Conclusions

From a broader perspective, these results provide an alternative pathway by which nano-structuring can influence material properties. Typically, a superlattice response is thought to arise either through localized or coherent effects. The latter concerns the coherence length of the states with respect to interface periodicity, while for localization, discrete confined quantum states exist that are different than those in the bulk. Neither of these views explain the changes in the vibrational response observed here, because they neglect underlying symmetry changes that can propagate into the constituent materials. When scaling the phases that constitute the material to unit-cell dimensions, the solid takes on a new symmetry that cannot be explained by a combination of the constituent materials. In these STO-CTO superlattices, this new structure results from octahedral coupling between the layers. Here, we have directly imaged these localized changes in symmetry and their impact on vibrations using a combination of STEM iDPC and monochromated EELS, with conclusions drawn supported by DFT. We further demonstrated how the observed localized phenomena evolve from locally impacting the superlattice at larger periods to dictating the global response of the superlattice as the period decreases via non-linear optical and phonon lifetime measurements. It is important to note that the reported changes in symmetry are not from the global periodicity of the superlattice. Rather, it is the local symmetry changes at the interfaces, and their spatial distribution, that ultimately dictate the entire macroscopic response of the solid as the period thickness decreases. Therefore, tailoring interfaces, and knowing their local response, provides a means of pursuing "designer" solids having emergent infrared and thermal responses not inherent within either of the constituent bulk materials.

# 5 Methods

## 5.1 Thin-film growth

Superlattices grown on NdGaO$_3$ substrates were synthesized using reflection high-energy electron diffraction assisted pulsed laser deposition. Further discussion of growth is found in reference 10 and its supplementary information. Superlattices grown on LSAT substrates were realized via molecular beam epitaxy as outlined in reference.[55]

## 5.2 Electron Microscopy

Scanning convergent-beam electron diffraction, ADF and iDPC images were acquired on a Thermo Fisher Themis Z-STEM operating at 300 kV. ADF and iDPC images were acquired with a 30 mrad convergence angle, 200 pA probe current, 625 nm/px dwell time. A 145 mm camera length projected onto the ADF detector with a 200 mrad outer radius and 40 mrad inner radius. The segmented ADF detector used for iDPC had a 38 mrad outer radius and 10 mrad inner radius. The position of metal sites was refined by thresholding, finding the center-of-mass, then fitting with 2D-Gaussians. The spacing of oxygen columns necessitated locating atomic columns manually.

Vibrational EELS spectra were acquired at 100 keV using a Nion HERMES monochromated aberration-corrected dedicated STEM with a convergence angle of 32 mrad, 25 mrad entrance aperture collection angle, and 0.413 meV/channel dispersions for the SL2 and SL4 acquisitions, and 0.826 meV/channel for the SL27 acquisition. In this paper, all EELS are acquired in an off-axis mode, obtained by shifting the electron diffraction pattern with respect to EELS entrance aperture. In EELS, delocalized dipole scattering can dominate signals and mask local variations in phonon populations which can be detected by impact scattering. However, by acquiring EELS only from electrons scattered out to high angles, the dipole scattering is reduced more than the impact scattering and the localized signals can be retrieved.[35,37] Here, we displace the optic axis from the EELS entrance aperture by ~50 mrad in the non-dispersive axis of the spectrometer, and only integrate pixels in the top half of the acquired signal. Thus, the off-axis EELS shown in this manuscript has an effective collection semi-angle of ~12.5 mrad that is scattered ~55.5 mrad from the central optic axis. By integrating the signal at higher angles, we preferentially select vibrations that have undergone impact scattering, which is a more spatially localized signal, and exclude the electrons that have undergone dipole scattering to low-angles, which is spatially delocalized.

Vibrational EELS background removal with fitted functions can introduce error because the realistic background does not have a functional shape, due to overlapping of the zero-loss peak and real spectral features such as non-resolvable acoustic or low-energy optic phonons, Therefore, we choose to take an alternate approach normalizing the spectra by multiplying by $E^2$ thus uniformly normalizing the spectra to a quadratic background. To increase signal-to-noise and compare with theoretical predictions we take the average of all spectra in STO, CTO, and interface layers, akin to layer DOS in DFT. The interface signal is defined as one unit-cell in length for consistency with structural characterization. The layer averaged signals from each layer can then be easily compared to one another and the average superlattice signal. In SL2, structural characterization showed that we cannot define a structural interface or structurally unique layers. We therefore had three choices; 1) define the entire period as a layer, which is the same as the average superlattice signal and provides no comparable spectra, 2) revert to the definition for chemically defined interfaces, which provides STO and CTO layer average spectra for comparison with each other and the average superlattice signal, or 3) use the structurally diffuse interface width of one unit-cell, which provides the interface, STO, and CTO layer. The third definition does not leave an actual STO or CTO layer, because only a single atomic plane of $TiO_2$ remains between interfaces. The lack of a complete STO and CTO layer was part of the rational for a single phase in the iDPC analysis, making choice (3) inconsistent with the structural analysis. Choice (2) would be inconsistent with the EELS analysis of SL27 and SL2. We choose choice (3) so that the EELS analysis between the three superlattices was consistent, and because atomic resolution conditions were not used in the EELS experiments making the one atomic plane delineation of layers infeasible. The lack of spectral change from STO to interface to CTO layer observed in the EELS analysis of SL2 then shows that the layers beave in a similarly, which is consistent with each having a similar symmetry.

**Calculations**

The DFT calculations used the Vienna ab initio Simulation Package (VASP) [56] with the projected-augmented wave (PAW) [57,58] method and the local-density-approximation (LDA).[59] Phonon calculations were performed using the LDA for exchange-correlation because it has been found to do better for phonons at the $\Gamma$ point, which is of interest here, in bulk CTO and STO.[51,60,61] The plane-wave basis energy cut off is 600 eV. The superlattice structural models were constructed by alternatively combining *Pbnm*-phase STO and CTO in the *c* direction with specific thicknesses. A SL8 model was chosen to obtain tilt-angles for a large-period superlattice knowing the interface coupling is limited to a few atomic planes, and the prohibitive computational requirements of simulating SL27. For structural relaxation, the structures were relaxed until the atomic forces were less than 0.01 eV/Å. The lattice parameters were also optimized for each superlattice model. Phonon calculations were performed using the finite-difference method. For structural relaxation and phonon calculations, the k-samplings are 6×6×6 for bulk STO and CTO, 4×4×2 for SL2 and SL4, and 2×2×2 for SL8, respectively. A full width at half maximum (FWHM) of 16 cm$^{-1}$ was used to plot the projected phonon density of states.

The phonon DOS of each model is obtained by performing weighted average over the respective constituent layers. The projected phonon DOS of a respective constituent layers is normalized by the number of atoms per layer to provide a consistent comparison between the three superlattices. The total phonon DOS of each model is then obtained by $n_{total} = (x \times n_{STO} + y \times n_{CTO} + z \times n_{int.})/(x + y + z)$, where x, y, and z are the numbers of atoms considered in each layer.

In particular, the phonon DOS of SL27 is obtained by averaging phonon modes of intrinsic bulk STO, intrinsic bulk CTO, and SL8 interface. The interface in all models is defined as one unit-cell on either side of the chemically defined interface, which is consistent with experimental and calculated structures. Since the primary structural changes are associated with $TiO_6$ octahedra, we can assume that the distinct vibrational state of different superlattices are primarily contributed by Ti/O-related vibrational modes. Therefore, we project the phonon DOS on Ti and O atoms, emphasizing the symmetry-phonon relation. Because only the phonon modes parallel to fields are activated, the phonon DOS is also projected in the (110) plane that is perpendicular to the electron beam.

### 5.4 Optical Spectroscopies

Raman spectroscopy was performed on samples synthesized atop NGO with a Horiba LabRam Raman instrument employing a 325 nm laser focused using a 40X/0.5 NA objective. Laser powers were verified to be inconsequential to the results. At this wavelength, the skin depth for the exciting UV-light is 26 nm within STO while being >1 μm for CTO. Despite the transparency of CTO, all Raman examined films were 200 nm in thickness and thus contain at least 100 nm of STO. This is more than 3 times the skin depth and thus the underlying NGO does not impact the Raman experiment. The monolithic samples exhibit a response expected from their bulk form. [62–66] Raman and FTIR spectra were fitted using a least-squares minimized linear combination of acquired monolithic spectra. This extenuated differences between the superlattice and constituent materials and helped remove the substrate response in the FTIR.

### 5.5 Second-Harmonic-Generation (SHG)

SHG measurements were performed on SL27, 6, 4, 3, 2, and 1 with nominal thicknesses of 200 nm atop NGO substrates. The home-built SHG microscope is centered on a 1040 nm Nd:YVO$_4$,

~100 fs Gaussian laser source that is focused to the sample surface at an incident angle of 45 degrees relative to the surface normal using a 10x microscope objective (NA = 0.28). The incident beam polarization is rotated using a half-wave plate. The forward-scattered beam, containing both the fundamental and second-harmonic frequencies, is collected with a lens. Through a series of band-pass filters, non-SH components are filtered out, while the second-harmonic component is focused to an amplified avalanche photodiode. The generated voltage is further amplified via lock-in detection demodulated at the laser repetition rate. The reported second-harmonic values are the parabolic coefficient determined by fitting the measured SHG intensity (e.g., lock-in photodiode response) as a function of incident laser power for each sample. The square-dependence in measured intensity vs. incident field indicates no higher harmonics are measured or optical leakage of the fundamental frequency is reaching the detector.

### 5.6 Time-domain Brillouin Scattering (TDBS)

The TDBS measurements were performed using an 80 MHz, 800 nm Ti:Sapphire oscillator (~100 fs pulses) that is split into two optical paths prior to reaching the sample. The first beam is used as a high-energy pump pulse, that, when focused to the sample surface, stimulates coherent acoustic phonon modes via rapid thermal expansion of the material. This pump pulse is frequency-doubled (400 nm) for these measurements to increase optical absorption in the STO-CTO layers. The second beam is sent down a mechanical delay stage to vary the time of which the pulse reaches the sample surface; this low energy 'probe' pulse monitors changes in the optical properties of the sample following excitation as a function of time-delay between the two pulses. As the coherent longitudinal phonon mode propagates through the superlattice, the probe beam partially reflects off the sample surface and partially off the coherent wave. The distance between these partial reflections evolves in time due to propagation of the phonon mode, and thus operates as a Fabry-Perot interferometer, where for distances that are integer multiples of the probe wavelength, constructive interference is observed in the signal, and for half-integer wavelength distances, the two reflections destructively interfere and reduce the signal. The temporal decay of these sinusoidally-varying oscillations is a direct monitor of the lifetime of the pump-generated longitudinal vibrational mode within the SL structure.

## Acknowledgements

ERH and PEH appreciate support from the Office of Naval Research through a MURI Program, Grant Number N00014-18-1-2429. Theory at Vanderbilt University was supported by the U.S. Department of Energy, Office of Science, Basic Energy Sciences, Materials Science and Engineering Directorate grant No. DE-FG02-09ER46554 and by the McMinn Endowment. Calculations were performed at the National Energy Research Scientific Computing Center (NERSC), a U.S. Department of Energy Office of Science User Facility located at Lawrence Berkeley National Laboratory, operated under Contract No. DE-AC02-05CH11231. The oxide heteroepitaxy synthesis work at Berkeley is supported by the Quantum Materials program from the DOE Office of Science, Basic Energy Sciences under Contract No. DE-AC02-05CH11231. R.R. also acknowledges the ARO MURI under agreement W911NF-21-2-0162.

## Author Contributions

E.R.H., J.A.H., and J.M.H. contributed to the acquisition, analysis, and understanding of all scanning electron microscopy data. D.B., A.O., and S.T.P contributed all density-functional-theory calculations. Z.P. and T.E.B. contributed acquisition, analysis, and understanding of UV-Raman data. J.R.M, T.E.B, and J.D.C. contributed acquisition, analysis, and understanding of Fourier-transform Infrared spectroscopy data. A.K.Y., R.C.H., R.E.H, J.R., and R.R. contributed growth expertise and samples used in the analysis. J.F.I and P.E.H. contributed an understanding of how the crystal and vibrational structure impacted broader material properties.

## Data Availability

The datasets generated during and/or analyzed during the current study are available from the corresponding author on reasonable request.

## Supplemental Information Figure Captions

# S1 Supplemental Information: Diffraction

The discussion of SADPs shown in Figure 1(b-d) of the main text detailed changes in the orientation of the *Pbnm* unit-cell that results from octahedral tilting, and further implicates that the octahedral tilting might also be what mediates the changes in orientation. In this supplemental section, simulations are presented that support and clarify ordered reflections in the SADPs, lattice parameters for the superlattices are quantified, and other features in the diffraction pattern that may implicate changes in structure or phonon softening will be discussed further. Lastly, scanning convergent-beam electron diffraction is used to spatially map octahedral tilting as a function of position, which is like the SADPs but includes spatial dimensions in addition to scattering dimensions.

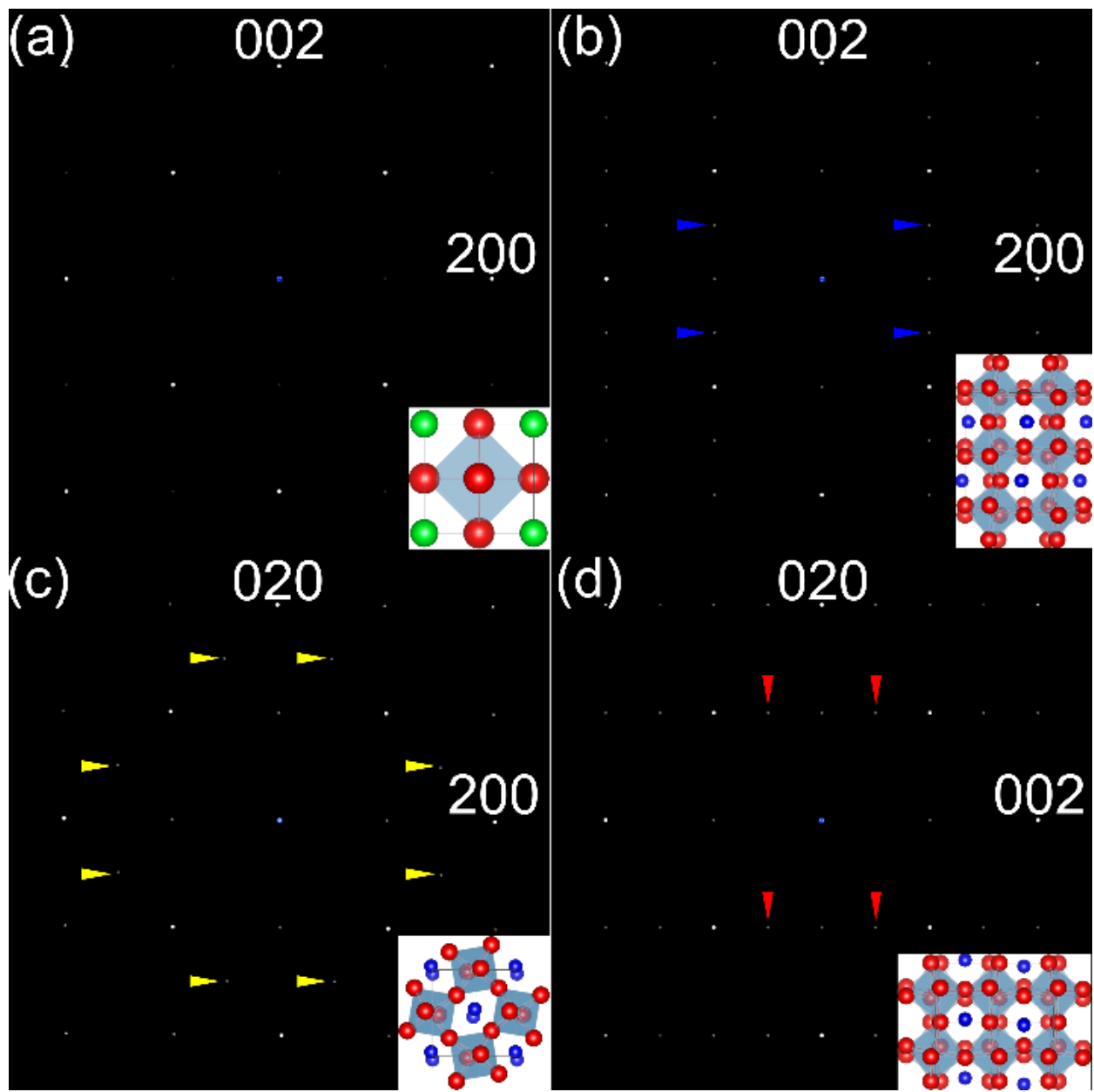

Figure S1. Simulated SADP for: (a) $Pm\bar{3}m$ and (b,d) $Pbnm$ along the out-of-phase tilt axis and (c) along the in-phase tilt axes. Blue, red, and yellow arrows indicate the same ½{201}$_{pc}$, ½{021}$_{pc}$, and ½{130}$_{pc}$ ordered reflections, respectively, observed in the experimental SADP shown in Figure 1. Note that here the ordered reflections have different Miller indices that are based on the $Pbnm$ pseudo-cubic axes while in Figure 1 the ordered reflections are labeled based on the STO axes aligned with the film geometry.

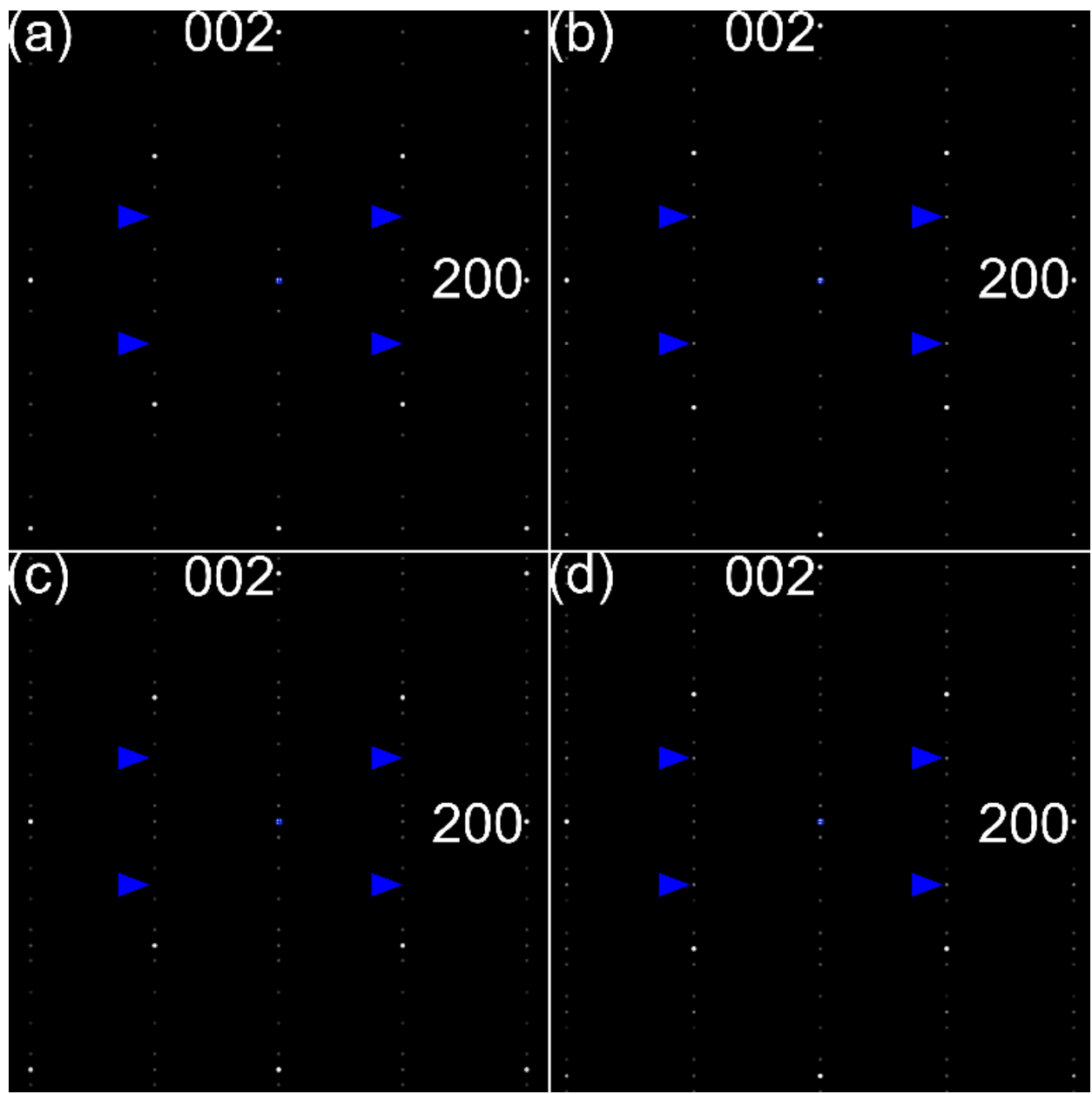

Figure S2. Simulated SADP for SL2 constructed form (a) four $Pm\bar{3}m$ and (b) two *Pbnm* unit-cells and SL4 constructed from (c) eight $Pm\bar{3}m$ and (d) four *Pbnm* unit-cells. Blue arrows indicate the absence of ordered reflections in $Pm\bar{3}m$ diffraction patterns and presence of ½{201}$_{pc}$ type ordered reflection in *Pbnm* diffraction patterns.

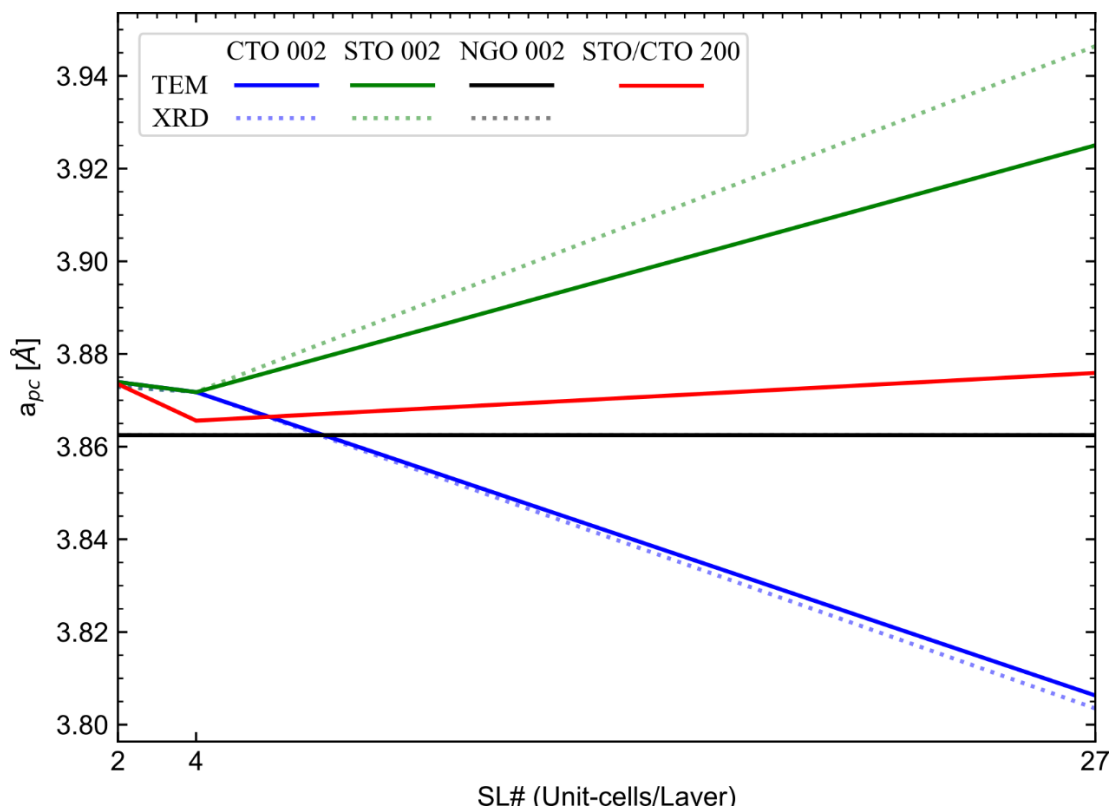

Figure S3. Calibrated lattice parameter calculated from the (002)$_{pc}$ Bragg peak using TEM SADP (solid) and x-ray diffraction (dotted) for CTO (blue), STO (green), and the NGO substrate (black). The in-plane lattice parameter for STO and CTO calculated from the (200)$_{pc}$ Bragg peak in a SADP is shown (red).

XRD was performed with a 2$\theta$-$\omega$ scan geometry to determine the out-of-plane pseudo-cubic lattice constant. The NGO substrates (002)$_{pc}$ was used for alignment and the calculated lattice parameter showed minimal deviation between samples. It is assumed that the measured lattice parameter of the NGO substrate should be identical for all techniques and all samples. Therefore, all SADP and XRD measurements of STO, CTO, and NGO were scaled such that the NGO (002)$_{pc}$ lattice parameter equaled the average NGO (002)$_{pc}$ lattice parameter of the XRD scans. Lattice parameters were extracted from the SADP in Figure 1, and the scaled values are shown in Table S1. The lattice parameters calculated from XRD experiments are shown in Table S2. The scaled lattice parameters from SADP and XRD are plotted together as a function of SL period in Figure S3.

Table S1. Scaled lattice parameters calculated from SADP in SL2, SL4, and SL27. Values are in Å.

| SL# | 002 | | NGO | 200 |
| --- | --- | --- | --- | --- |
| | STO | CTO | | STO/CTO |
| 2 | 3.874 | | | 3.873 |
| 4 | 3.872 | | 3.862 | 3.866 |
| 27 | 3.925 | 3.806 | | 3.876 |

Table S2. Scaled lattice parameters calculated from XRD in SL2, SL4, and SL27. Values are in Å.

| SL# | STO | CTO | NGO |
| --- | --- | --- | --- |
| 2 | 3.873 | | |
| 4 | 3.872 | | 3.862 |
| 27 | 3.804 | 3.946 | |

From the measured lattice parameters, it was found that the out-of-plane lattice parameter converged to a single STO-CTO value that was approximately the average of the two unrelaxed lattice parameters of SL27. This is the general theme observed in all crystallographic and vibrational structure analysis herein, that is, as the number of unit-cells in a layer decreases the structure converges to a singular, uniform, intermediate structure. The in-plane lattice parameter is different (but nearly equivalent) to the substrate in SL4 and 27. SL2 had in-plane lattice parameters like their out-of-plane lattice parameter and were larger than SL 4 and 27, indicating that the film in-plane lattice parameters are not determined by the substrate.

In the SADPs shown in Figure 1(b-d) a rich set of information is present between the Bragg peaks.[67] In SL27 shown in Figure 1(b), closely spaced superlattice reflections extend from the Bragg peaks making them look streaked. The same superlattice reflection are seen in SL4 and SL2, but are much further spaced and appear as distinct peaks because of the much larger real-space periodicities. An interesting diffuse background intensity is found in SL4 at the location of the ordered reflections. We will speculate to their origin but defer to future research to ascertain the true origin of the diffuse intensity. Diffuse intensity in the background of diffraction patterns is associated with the loss of long rand coherence, whether it be from structural static disorder or thermal vibrations. In the present case, we find that the superlattices contain a high degree of long-range static ordering, as clear from the sharp fundamental, ordered, and superlattice Bragg reflections. This remaining option is thermal vibrations. In many phase transitions a phonon mode at a specific momentum vector causes the transition to a lower symmetry structure. In these cases the thermal diffuse scattering, typically considered as a uniform Gaussian like distribution centered at $q$=(000), can become non-uniform and slowly develop into Bragg peaks as the mode softens and forms new Brillouin zones. [68–70] In the case of SL4 we observe with iDPC that octahedral tilting is present throughout the entire superlattice structure, including within STO layers where no natural tilts are present at ambient conditions. SL4 also has the smallest number of unit-cells per layer where the layers can be well defined, see section S2 and the main text for elaboration. From the continuous change in tilting, it is expected that the phonon associated with the tilting is neither completely hardened or softened through the entirety of the structure and may therefore produce a non-uniform thermal diffuse background.

Convergent-beam electron diffraction patterns were acquired for each pixel of a line scan, known as scanning convergent-beam electron diffraction (SCBED), across the STO-CTO interfaces. Ordered Laue zones in the atomic resolution SCBED allows mapping of octahedral tilt. [20] Line scans of a two, four, and twenty-seven unit-cell superlattice are shown in Figure S4.

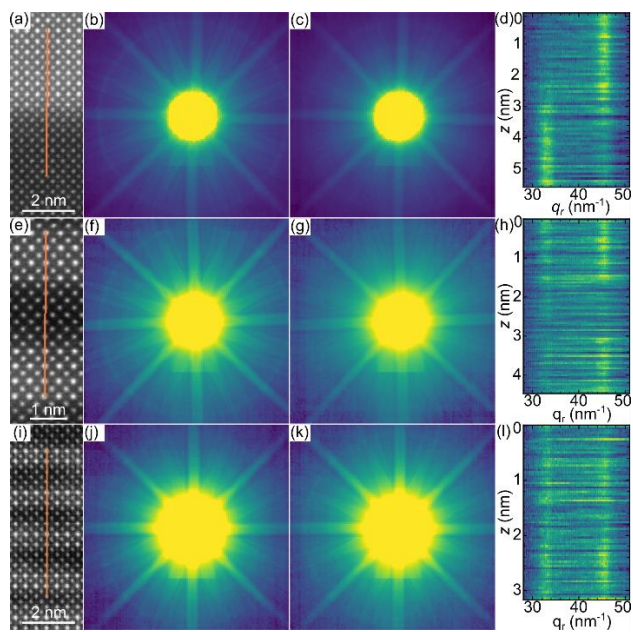

Figure S4. Scanning-CBED measurements from (a-d) SL27, (e-h) SL4, and (i-l) SL2. The position of line scans is indicated in the (a,e,i) ADF reference image. The PACBED pattern from the STO and CTO layers are shown in (b,f,j) and (c,g,k), respectively, show layer dependent fundamental and ordered Laue zones. (d,h,l) Background subtracted radially integrate CBED patterns show the existence of ordered reflections, indicative of octahedral tilting, as a function of position as the probe scanned across the superlattices.

The abrupt change in contrast of A-site atoms in the high-angle annular dark-field ((HA)ADF) signal shows the chemical abruptness of the superlattices. To visualize the relative difference in diffraction patterns taken from STO and CTO layers we compare the position averaged CBED (PACBED) from each layer, as shown in Figure S4(b,c). In the large period twenty-seven unit-cell superlattice the STO PACBED had a Laue zone corresponding to the intersection of the Ewald sphere with fundamental Bragg reflections of the $Pm\bar{3}m$ reciprocal lattice. The CTO PACBED had an identical fundamental Laue zone and an additional Laue zone, with half the radius of the fundamental, corresponding to the intersection of the Ewald sphere with half-integer ordered reflections appearing from $TiO_6$ tilting. The existence of tilting in CTO and not in STO further supports the conclusion from SAED that STO and CTO are relaxed to their monolithic phases. When the number of unit-cells in a layer is reduced to four per layer the intensity of fundamental and ordered Laue zones in the CTO PACBED were much more similar, as shown in Figure S4(g). This could indicate some reduction in the octahedral tilt-angle, but in general shows that octahedral tilting still exists in CTO as expected from the SAED. With the decrease in layer thickness the STO PACBED now shows both a fundamental and ordered Laue zone much like CTO, as shown in Figure S4(f). Therefore, we conclude that the STO has inherited $TiO_6$ tilting from the CTO layers because of the reduced layer thickness. The inheritance of octahedral tilting persist with a further reduction to two unit-cells per layer and the comparable intensity of ordered and fundamental Laue zones is seen in both layers. The observed ordered reflections show that the STO layers are inheriting octahedral tilting and becoming adapting to the CTO crystal structure.

We know that the layers are changing and wanted to investigate more local changes. By radially integrating the CBED and subtracting a background power law the relative intensity of ordered and fundamental reflections at each pixel of a line scan is compared, as shown in Figure S4(d,h,l). In SL27 the intensity of the fundamental Laue zone is uniform in both CTO and STO layers and the intensity of the ordered Laue zone is uniform in the CTO layer showing that the structure of the layers is uniform, as shown in Figure S4(d). At the STO-CTO interface a transition region occurs over three atomic planes. In this transition region an ordered Laue zone appears in STO and gradually increases intensity as the probe entered the CTO. There is therefore a chemically abrupt and structurally diffuse interface that results from the coupling of STO and CTO layers, much like $La_{0.5}Sr_{0.5}TiO_3$/CTO and $NdGaO_3$/CTO interfaces.[23] When the number of unit-cells is decreased to four or two unit-cells per layer, where the structurally defined interface size is the same size as the chemically defined layers and octahedral tilting is present everywhere, the distinction between an interface region becomes less discernable. This could, in-part, be from relatively small contrast changes buried in noise, but implicates that the changes in ordered Laue zone intensity, indicating inheritance of octahedral tilting in STO and reduction of titling in CTO is from the overlapping structurally diffuse interfaces and the superlattice is transitioning to a system structurally defined by such interfaces.

## S2 Supplemental Information: ADF and iDPC

In the STO layer of SL27 an in-phase tilt-angle of 1.84° and 2.69° STO was measured in the $TiO_2$ and AO planes, respectively, since no tilting is present and represents a bound to measurement error. In CTO out-of-phase tilt-angles of 10.04° and 10.32° in the $TiO_2$ and AO planes, respectively. In SL4 the quantified out-of-phase tilt-angle profile was sinusoidal with an average of 6.34° and 7.39° in the $TiO_2$ and AO planes, respectively, demonstrating that the layers tilts have accommodated to approach the mean or interface value observed in the larger-period SL27. In SL2 the tilt profile did not have any systematic tilt-angle related to the layer periodicity appearing "dephased" with a nearly constant tilt-angles of 7.138° and 6.096° in the AO and $TiO_2$ planes, respectively.

The relative proportion of layers and interfaces can be quantitatively assessed by comparing the volume fraction of interfaces, STO, and CTO defined by

$$V_{int} = \frac{1}{n-1} \tag{1a}$$

$$V_{STO} = V_{CTO} = \frac{n-2}{2(n-1)} \tag{1b}$$

In a large-period superlattice $n$ is large so $V_{ATiO_3} \gg V_{Int}$. For a four unit-cell superlattice $V_{ATiO_3} = V_{Int}$ such that the interface is of equal importance to still existing layers. In the extreme of the two unit-cell superlattice $V_{ATiO_3}=0$ and $V_{Int}=1$ such that the entire superlattice is an ordered structure.

## S3  Supplemental Information: DFT

In SL8 there is an asymmetry in tilt-angle gradient at the interface. The coupling extends further into the STO layer than into the CTO layer. The STO tilt-angle is also non-zero, unlike in bulk STO. The preferential increase in STO tilt-angles, relative to decreasing CTO angles, suggests that the O-Ti-O tilt mode is softer in STO. We cannot entirely rule out the influence of strain on this behavior.

Relaxed structures have lattice parameters different than the bulk STO and CTO lattice parameters. In SL8 the structure did not relax to the bulk like lattice parameters in the CTO and STO layers. A 1.8% compressive strain and 1.7% tensile strain is present in the STO and CTO layers, respectively, relative to bulk like phases.

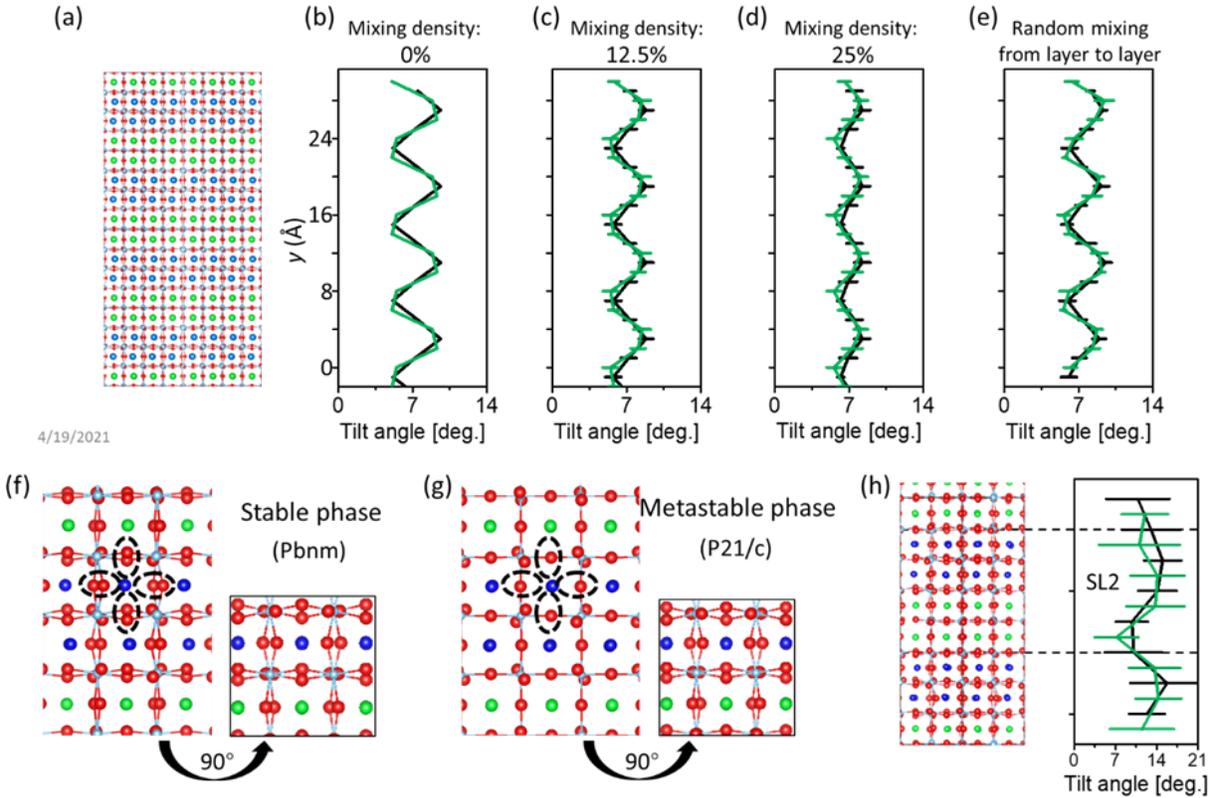

Figure S5. (a) Atomic model of SL2. (b-d) O-Ti-O tilt-angles as the uniform intermixing density changes from 0% to 25%. The y-axis scales are the same as in (a). ). (e) O-Ti-O tilt-angles when a 14.1% intermixing density is distributed nonuniformly in the layers. (f) Stable *Pbnm* phase of SL2. (g) Metastable phase of SL2 with *P21/c* symmetry. (h) O-Ti-O tilt-angles when oxygen vacancies are present, distributed randomly in the layers, with an overall density of $2.2 \times 10^{21}$ cm$^{-3}$ (this is a high density of vacancies, limited by computational costs)

Tilt-angles with different intermixing densities are calculated and shown in Figure S5(b-e). As the uniform intermixing density goes up, the tilt-angles are greatly decreased to an average value of 7°. Meanwhile, if the intermixing density is nonuniform (which is more likely experimentally), the tilt-angles not only decrease but also show a dephasing (Figure S5(e)). The narrowing of the oscillation that is caused by intermixing in Figure S5 is an indication that intermixing and/or thermal effects that can occur at room temperature may be responsible for the absence of oscillations in the experimental data of Figure 1(m). DFT calculations predict that SL2 may have a novel metastable phase with *C2/m* symmetry. Comparing with the stable *P21/c* phase, the metastable phase brings tiny lattices change to the crystal, namely a compression of 0.35% in *a*, and a stretch of 0.35% and 0.1% in *b* and *c*, respectively, and an increased total energy of 1.85 meV/atom. The small energy increase and relationship between the structural space groups suggests that the existence of the metastable phase is highly possible, particularly at elevated temperatures. For the metastable phase, as shown in Figure S5(g), along the [110] direction the horizontal and vertical O-Ti-O tilt-angles are zero for the higher symmetry-phase SL2, while along the [1$\bar{1}$0] the tilts in the low-symmetry phase are still present. In such a case, measuring experimentally at a specific direction would obtain smaller tilt-angles than the expectation based on the stable state. The presence of rotational domains could lead to an averaging effect. All the above suggests that the presence of a temperature-dependent metastable phase and a small amount of intermixing may account for the constant angles in the experimental results of SL2. Moreover, oxygen vacancies are also considered in theoretical calculations. As shown in Figure S5(h), the existence of oxygen vacancies with a density of 2.2x $10^{21}$ cm$^{-3}$ cause the O-Ti-O tilt angles to increase significantly to over 14°, which is not the case with the experimental results. This result excludes the influence of oxygen vacancies in the experiments and confirms the high quality of the samples.

Additional discussion of Figure 3(b): Since the lattice parameters are optimized for the interfaces in the large-period calculations (e.g., SL8), there is always some strain on the STO and CTO layers. In our averaging method, we find very good agreement between the calculated SL27 phonon DOS and the experimentally measured EELS phonon spectra for SL27. By comparing the black curves with the orange curves in each model, we see that as the layer thickness decreases, the phonon DOS converges towards that of the interface curve, demonstrating the dominance of the interface in thinner layer superlattices. We also note that the mode energies of the three peaks discussed in Figure 3, have energies that correlate with Slater- and Axe-type displacements of the TiO$_6$ octahedra for a range of perovskites.[71] We also show a visualization of select modes in Figure S6 and Figure S8(b,c).

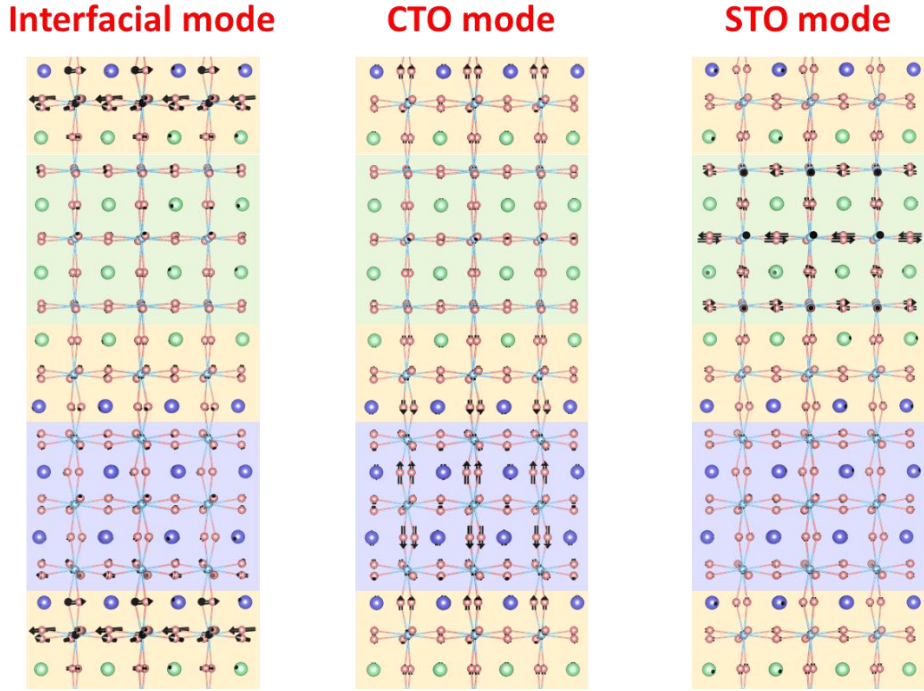

Figure S6. Eigenvectors for interface, CTO, and STO modes in SL4.

The eigenvectors for three typical modes ($q$=0), i.e., an interface mode, a CTO-layer mode, and a STO-layer mode are shown in Figure S6 to emphasize their spatial extent. The three phonon modes are located at 72.3, 90.9, and 70.8 meV. These modes were not selected to correspond to specific peaks in the projected phonon DOS, but to visualize the spatial localization of vibrations, which does not exist in uniform bulk materials. The vectors can be envisioned as displacement vectors at one moment of time during vibrations. We find that the interface modes are localized to the three-atomic-plane-wide structurally diffuse interface. The STO and CTO layer modes are likewise localized within the layers. We can therefore use DFT to visualize the localization observed in vibrational EELS.

The phonon DOS discussed in the main text is Ti and O projected. This was done to emphasize the impact of octahedral rotations on the phonon DOS and because the EELS analysis energy range was >30 meV. For completeness, we show the remaining A-site projected phonon DOS (PPDOS) in Figure S7. The Sr peak is lower energy than the Ca peak, because Sr is heavier. The Ca+Sr curve in SL2 and SL4 are very similar suggesting that their A-site vibrational response is similar. Most importantly for our justification to neglect A-site bonding, thus emphasizing the impact of octahedral rotation, is that all A-site vibrational responses are less than 30 meV. The A-site vibrations only provide a continuously decaying background to the DOS above 30 meV.

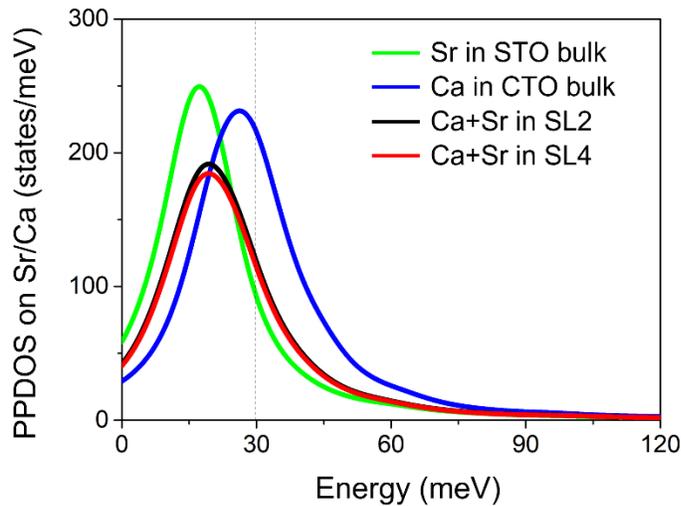

Figure S7. A-site projected phonon DOS.

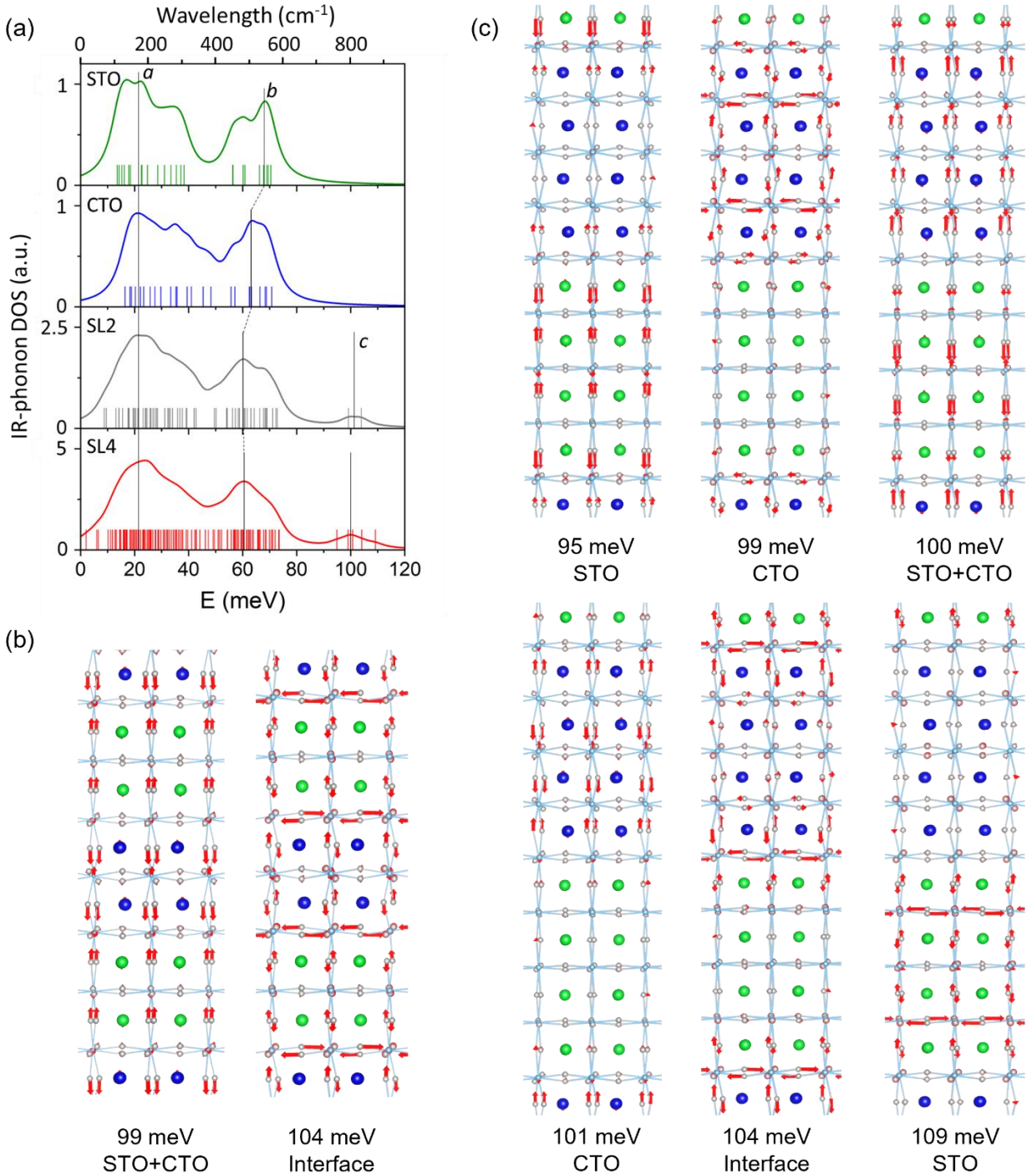

Figure S8. **Infrared active phonon DOS and eigenvectors showing emergent optically active vibrational response in superlattices.** (a) IR-phonon DOS SL4 and SL2 showing the emergence of IR active high-energy vibrations that are not active in either STO or CTO. Eigenvectors of vibrational displacements in (b) SL2 and (c) SL4 showing layer and interface IR active modes.

As a demonstration of the impact of the present work on understanding emergent properties that derive from structural and vibrational features of the SLs, we show in Figure S8(a) the predicted IR-active-phonon DOS for the bulk STO and CTO and the IR-active-phonon DOS for SL2 and SL4. The IR-active-phonon DOS can be compared to the experimental FTIR spectra shown in Figure 4, Figure S12, Figure S14, and Figure S15. We note emergent IR-active phonon modes at surprisingly high energies around 100 meV and also at very low energies. The emergent IR-active phonon modes consist of Ti-O modes localized within STO and CTO layers, delocalized between STO+CTO layers, and localized to interfaces. The density of the latter increases with increasing layer length. These emergent phonon modes are likely to underpin emergent properties, e.g., IR absorption and Raman spectra. Since in complex oxides, structure (e.g., octahedral tilts) and phonons are strongly coupled to electronic and magnetic properties, knowledge of the emergent phonons would help engineer novel properties. In particular, the displacement vectors of the high-energy modes have intriguing localization properties, either at the interfacial $TiO_2$ planes or within the layers that can potentially host unique spin structures.

## S4  Supplemental Information: Vibrational EELS

ADF signal was collected simultaneously with EELS data was used to identify the position of STO and CTO layers and their interfaces. This was done by finding the inflection points in ADF signals via differentiation of the signal, peak finding, then manually removing irrelevant positions. An example of interface assignment for an off-axis geometry acquired from SL27 is shown in Figure S9.

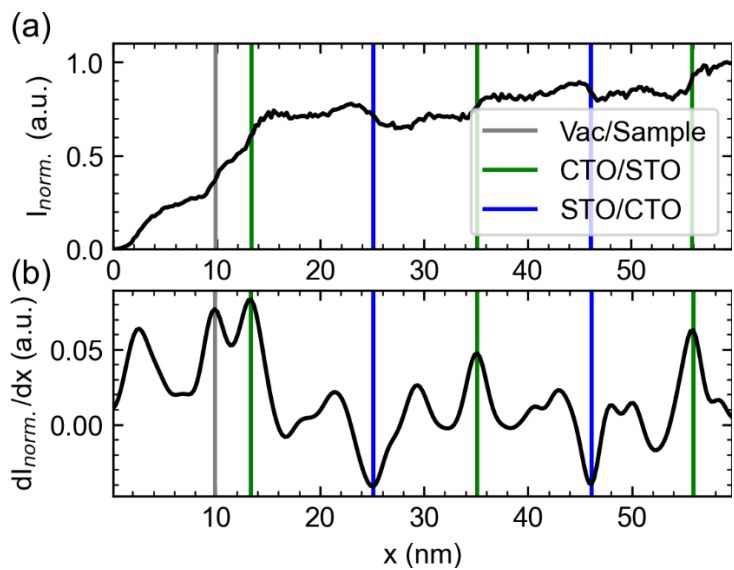

Figure S9. Example of interface assignment in SL27 off-axis signal.

In off-axis EELS line scans, ADF signals are asymmetric with Bragg reflections projecting onto the detector resulting in strain and diffraction contrast, which causes peaks in the ADF signal. An initial concern of qualitatively comparing the vibrational response of the layers and interfaces was that spatial differences in the differential scattering cross-section (with respect to **q**) could easily be confused with shifting peaks leading to a misinterpretation of localized interface modes. One way to determine if the apparent changes are from a change in scattering cross-section or peak energy is to directly compare the spectra. The signal in each layer was averaged into a position averaged energy-loss spectra (PAELS) to reduce the number of spectra to compare and aid interpretation, as shown in Figure S10.

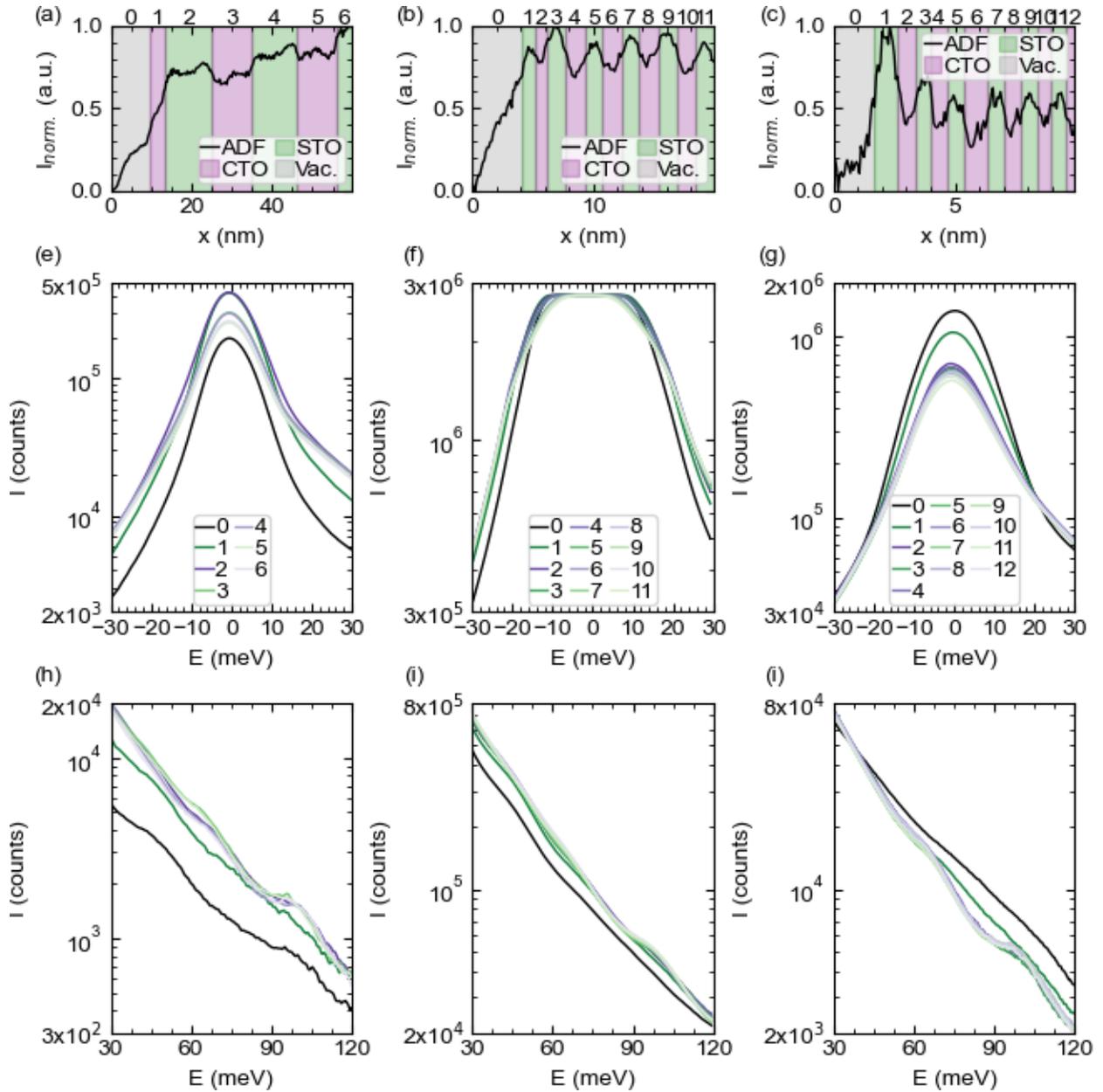

Figure S10. (a-c) ADF signal during the line scan with marker signifying defined layers in the off-axis geometry of (a,e,h) SL27, (b,f,i) SL4, and (c,g,i) SL2. PAELS of the (e-g) quasi-elastic peak and (h-i) loss region.

The quasi-elastic peak of the PAELS centered at $E=0$ meV is lowest when in vacuum for the off-axis geometry because localized quasi-elastic excitations are not excited. When the probe enters the material, phonons begin to inelastically scatter the incident electron and the quasi-elastic peak increases in intensity. One would expect that the quasi-elastic peak would then linearly increase with thickness like the ADF signal, which is in-part a result of impact scattered phonons. Instead, the quasi-elastic response decreased with each successive period and each layer in a period had nearly the same magnitude quasi-elastic peak. The exact reason for this occurrence is not known, but the overall decrease in signal could be attributed increase scattering probability of other inelastic excitations, such as plasmons and core-states. Quickly after the probe is in the material the tails of the quasi-elastic peak converge to similar values allowing for the intensity of peaks on the tail of the quasi-elastic peak to be compared directly without considering the influence of the total inelastic differential cross-section. This is further shown by the vibrational response intensity and energy difference between STO and CTO (for example see the ~65 meV region) and similarity between each CTO or STO layer. The convergence was conservatively set at layer 3, 5, and 5, for SL27, SL4, and SL2, respectively.

We then include a finite interface width into the PAELS, defined as one ±0.39 nm unit-cell, which is approximately one unit-cell on each side of the interface and contains the structurally diffuse interface measured in iDPC experiments. These newly defined layers and convergence point are used to form the layer average EELS in Figure 3 and Figure S11. Layer-averaged spectra in Figure 3 are shown as larger panels in Figure S11 to emphasize the small differences in peak energies, which are also listed in Table S3.

Table S3. Peak energies identified in the layer-averaged EELS spectra of Figure 3 and Figure S11. All energies are listed in meV.

|  | Peak 1 | | | Peak 2 | | | Peak 3 | | |
| --- | --- | --- | --- | --- | --- | --- | --- | --- | --- |
|  | STO | Int. | CTO | STO | Int. | CTO | STO | Int. | CTO |
| **SL27** | 40.1 | 45.5 | 46.2 | 66.7 | 66.1 | 63.3 | 100.3 | 98.6 | 96.8 |
| **SL4** | 45.3 | 45.9 | 45.9 | 69 | 68.2 | 67.3 | 98.2 | 96 | 95 |
| **SL2** | 36.2 | 37.3 | 38.5 | 64.3 | 63.5 | 63.1 | 98.8 | 98.1 | 98.8 |

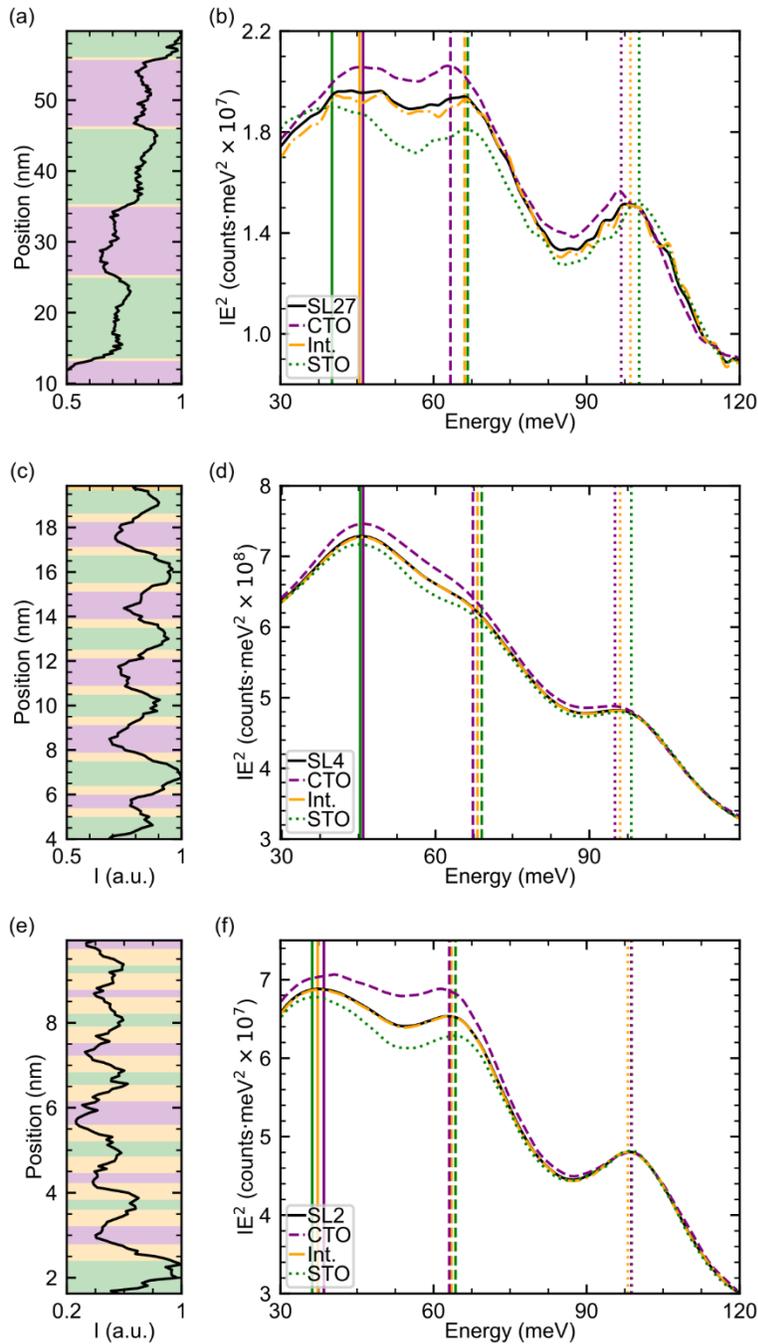

Figure S11. Monochromated STEM-EELS (b,d,f) layer-averaged spectra and (a,c,e) simultaneously acquired ADF signal.

In SL27, the STO, CTO, and interface layers each have unique responses that agree with trends found in DFT experiments. DFT informs that the vibrations in the 30-120 meV energy range are from the Ti and O sublattices, and that the peaks in the Ti+O projected DOS shift with tilt-angle.

We can conclude that the experimental observations are from layer-to-layer changes in $TiO_6$ tilt. To further emphasize this conclusion, the total and interface spectra can be compared. One would expect the interface spectra to match the total superlattice spectrum if unique interface vibrations are not present because both would be a linear combination of 50% STO and 50% CTO. Instead, there are discrepancies between the interface and total spectra that originate from vibration in the structurally diffuse interface. The small volume fraction of the region ascribed to the interface relative to the large STO and CTO layers suppresses the interface contribution to the total response. See section S2 for discussion regarding the volume fraction of layers. With the discrepancies, and agreement with DFT, we can conclude that the vibrational EELS experiments of SL27 is measuring local changes in vibrations at the interfaces that are a result of an octahedral coupling region. Furthermore, the interface and total spectra are nearly identical when the number of unit-cell per layer is reduced to four, such that the volume fraction of interfaces is identical to the volume fraction of either STO or CTO layers. Now that the interface represents an appreciable portion of the material, the interface vibrational response emerges and contributes appreciably to the total response. The interface contribution to the total vibrations is in addition to the influence that the interface has on the tilts in the STO layers, which will make the STO vibrational response more like the interface response, as also described by DFT.

It is useful to assess the current observations with regard to models for mixed systems, such as the virtual crystal model or two-ion model. [48,72] In the virtual crystal model structural or chemical heterogeneity at interface are incorporated into an interface layer, much like presented in the current work. [72] However, in the virtual crystal model, the interface layer is assigned properties that are effectively represent a disordered combination of the bounding materials. In the present case the interface is not disordered and contains vibrations that cannot be explained by the bounding materials. The structurally diffuse interface cannot be captured by a virtual crystal because of its unique structural and vibrational state. In a two-phonon model of superlattices phonons are grouped into either coherent or incoherent phonons. [48] The coherent phonons with long wavelengths and mean-free-paths can propagate through the material unscattered, while the incoherent phonons have short wavelengths and mean-free-paths and scatter from the interfaces. In SL27, the two-phonon model may adequately describe the transport of the material with the exception that the scattering probability of some incoherent phonons may not be as large as in a structurally and chemically abrupt interface because the interface structure and vibrations can mediate the transition from one layer to the other. In SL4, the incoherent phonons will also scatter form the interfaces, and there is a much higher density of interfaces so there will be a larger accumulation of scattering probability across the thickness of the superlattice. However, the incorporation of tilt into the STO layers makes the phonon modes more like the CTO modes such that transmission might be increased. In short-period case of SL2, the concept of a two-phonon model need not be considered. The structure in effect no longer has interfaces and the vibrational structure of the entire superlattice is uniform.

## S5 Supplemental Information: UV-Raman and FTIR

Complimentary UV-Raman and FTIR spectra were acquired to show the response as a function of interface density.

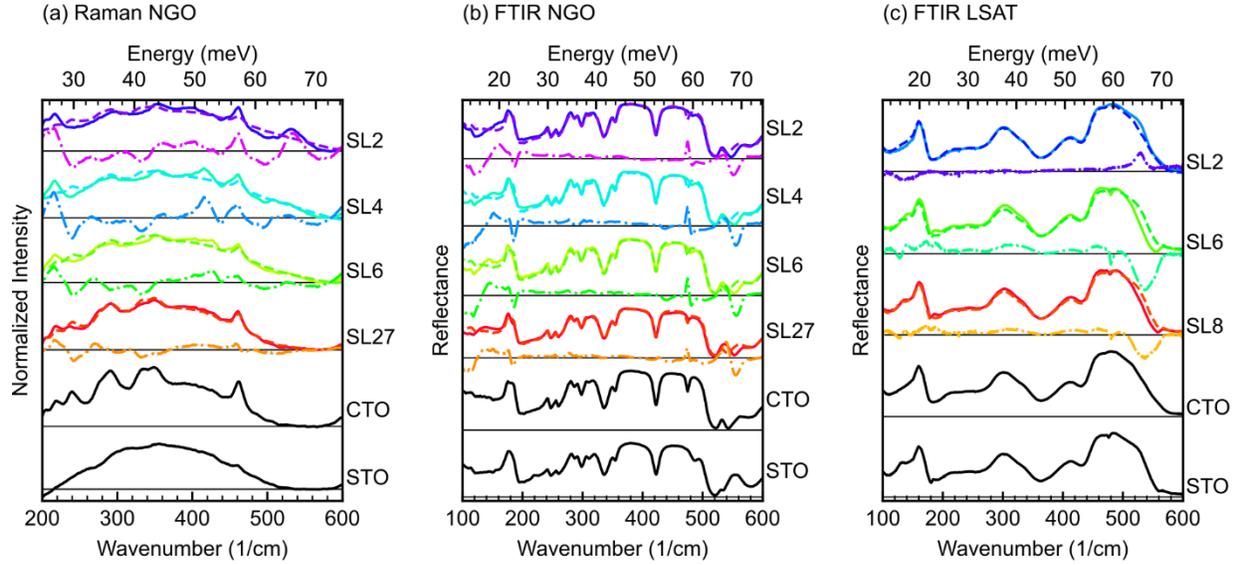

Figure S12. **Interface driven modifications in the macroscopic Raman and infrared response of STO-CTO superlattices.** Raw (solid), fitted (dashed), and residual (dot-dashed) data for (a) UV-Raman and (b) FTIR spectra taken from superlattices on an NGO substrate, and (c) FTIR spectra taken from superlattices on a LSAT substrate. 200 nm STO and CTO thin-films on substrates used to fit superlattice spectra are shown as the bottom of each panel. Difference curves are scaled by a factor of two for clarity.

To accentuate the changes that are unique to the superlattice structure, the residual from a fitted linear combination of STO and CTO on substrates is quantified, as shown by dot-dashed lines in Figure S12. With this approach, emergent Raman responses are observed near 220 cm$^{-1}$ (27 meV) and 450-550 cm$^{-1}$ (60-70 meV) that decrease intensity as the superlattice period is increased (Figure S12(a)). A similar evolution is observed in complementary FTIR measurements (Figure S12(b,c)). These emergent modes are similar in energy to that those observed on locally with EELS and DFT. We therefore conclude they are a consequence of the alternative symmetry existing at the interfaces of the superlattice. As such, their strength necessarily scales with interface density, resulting in the strongest response for the moderate- (SL4) and short-period (SL2) superlattices. Simply put, there are more interfaces or tilts characteristic of an interface in these films and thus the interfacial signal is stronger.

Below we present larger panels of spectra in Figure S12, as shown in Figure S13-13. We include vertical grey dotted lines in Figure S13 at 213, 448, 520, and 732 cm$^{-1}$ where strong first-order Raman features would appear if the substrate was being sampled.[73] Features in the region between 200-400 cm$^{-1}$ have previously been interpreted, for CTO, as resulting from O-Ti-O bending vibrations while features in the region between 400-600 have been assigned as Ti-O$_3$ torsional modes.[49,62,65] Due to the large mass of Ca or Sr atoms, vibrational modes involving their motion are not expected to be observed in the region between 200-500 cm$^{-1}$, but rather would occur at lower frequencies not observed by our experiments. Thus, although we cannot make a concrete assignment to each emergent Raman feature, we can infer that spectral features in this region reflect properties of the Ti-O sublattice.

To understand the presence of interface signal and expand upon the difference method described in the main text, we take the absolute value of the residuals then sum above 400 cm$^{-1}$ (49.6 meV) where the majority of observed changes occurred, as shown in Figure S16. The Raman residual decreases with increasing period thickness meaning that the scattering probability from interface modes decreases. The opposite is seen in the FTIR data, where minima describe more interaction with interface modes.

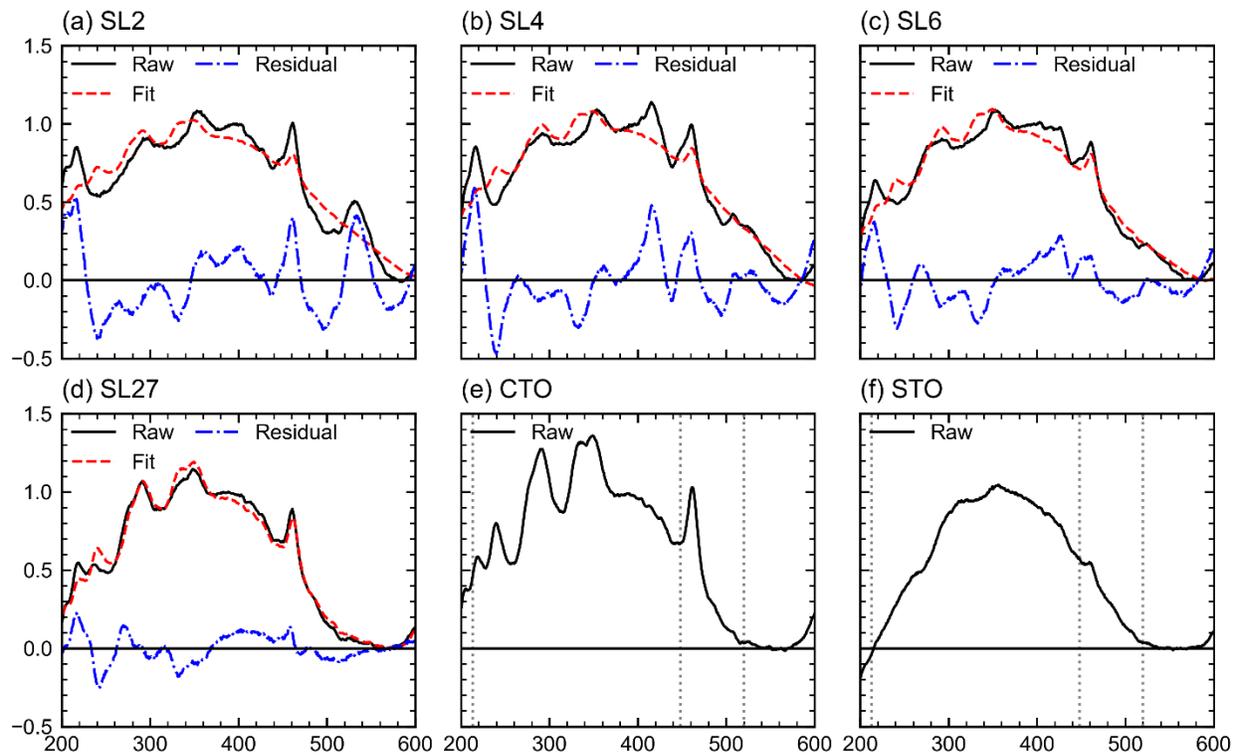

Figure S13. UV-Raman acquired from thin-films on NGO substrates. Grey dotted lines in (e) and (f) indicate energies where first-order NGO Raman peaks would appear if the substrate was being sampled.

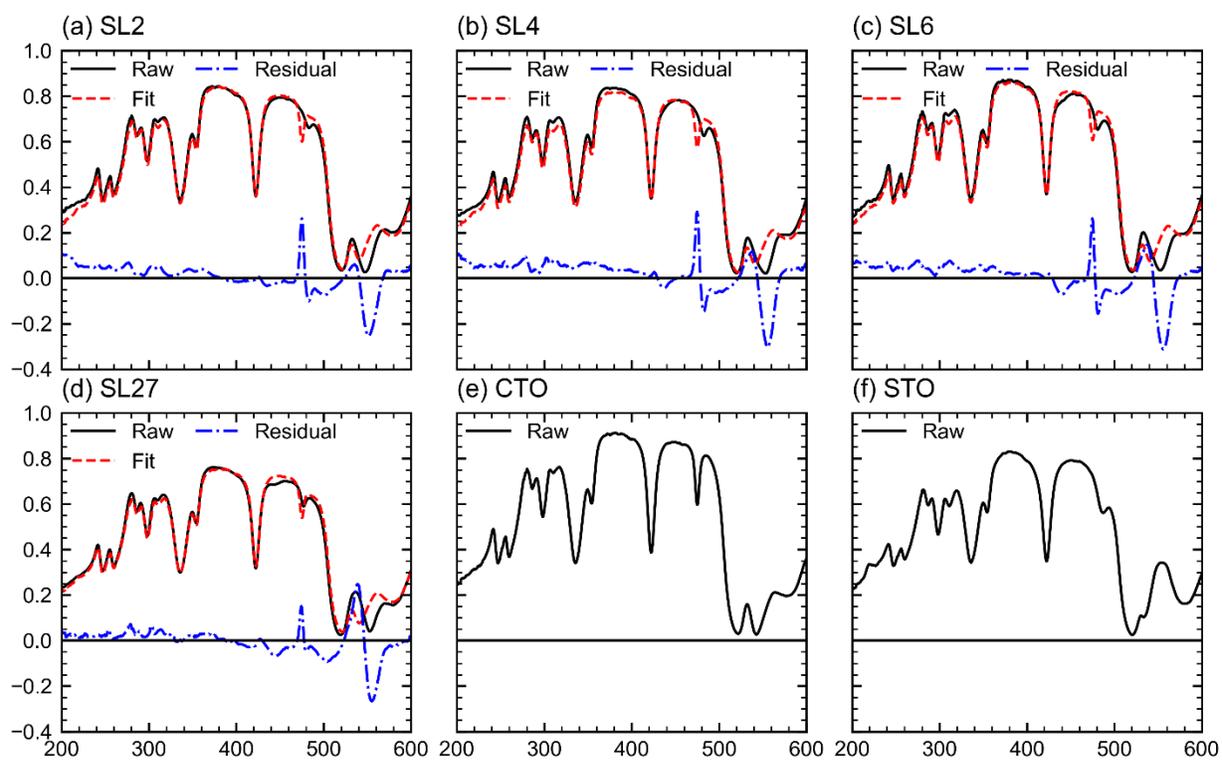

Figure S14. . FTIR acquired from thin-films on NGO.

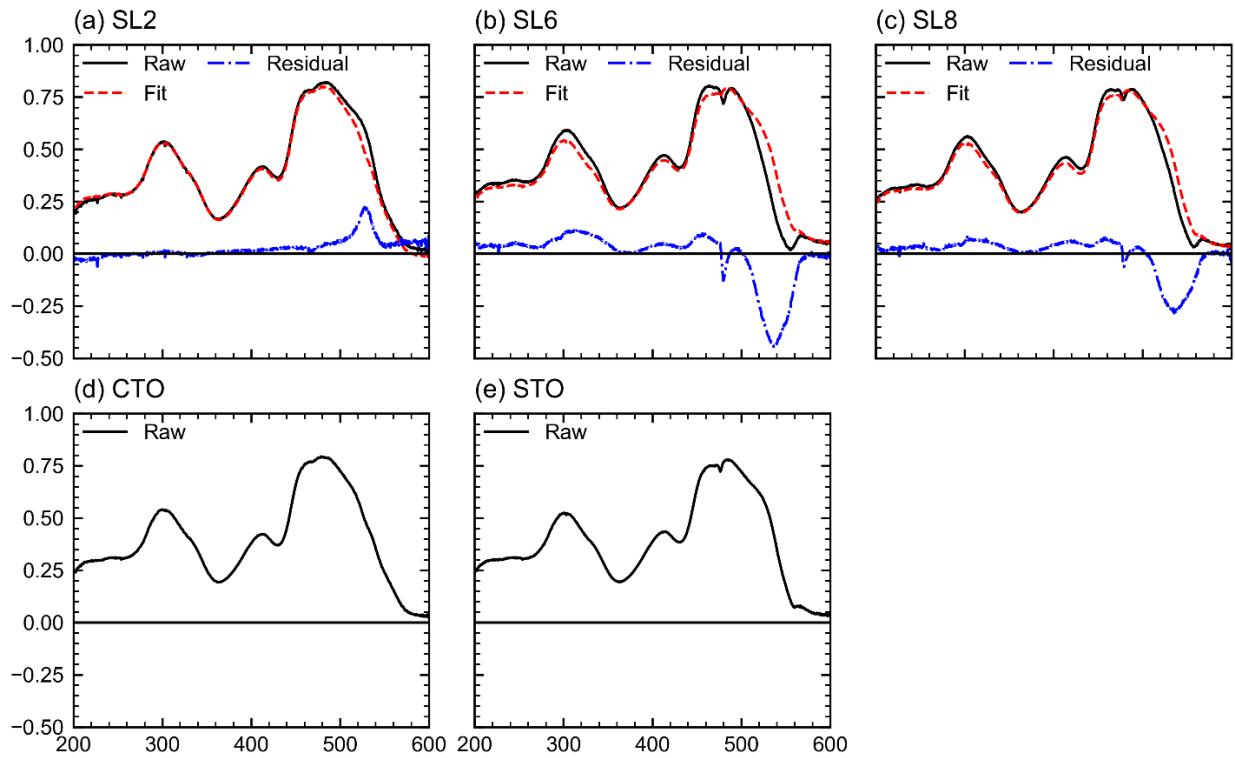

Figure S15. FTIR acquired from thin-films on LSAT.

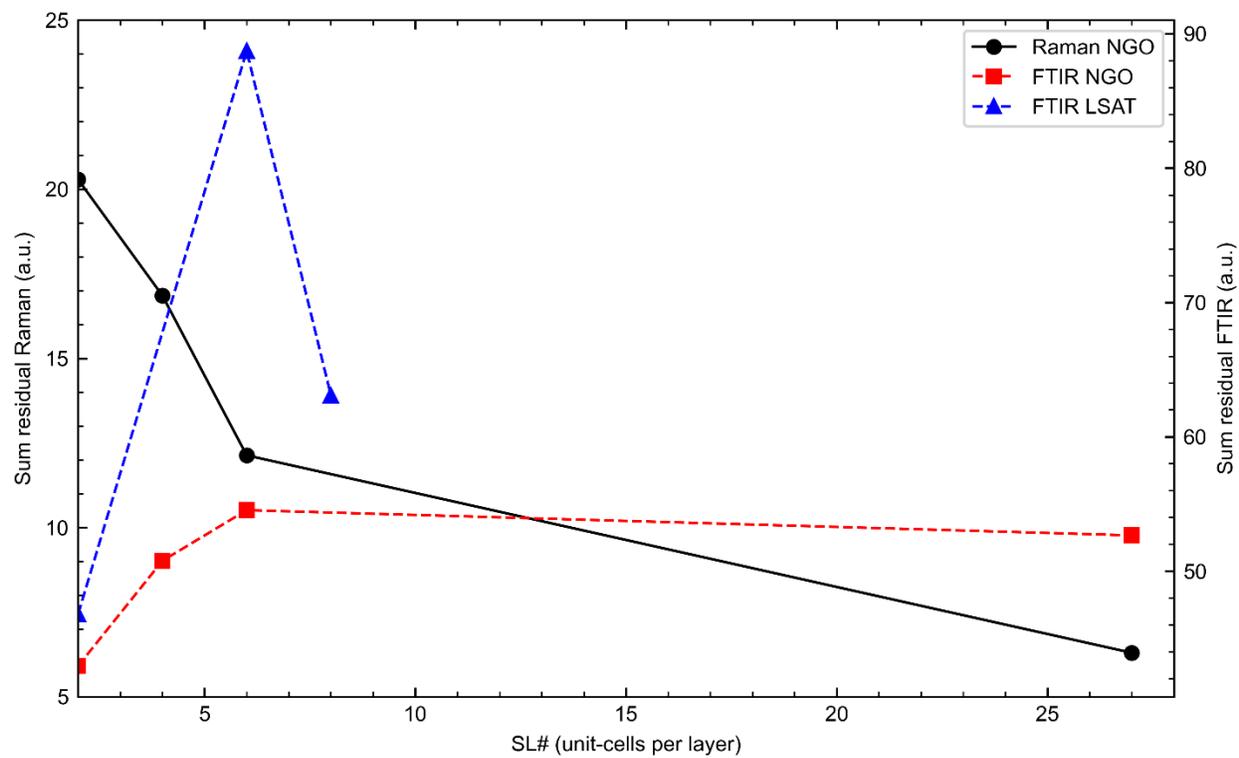

Figure S16. Sum of residuals above 400 cm$^{-1}$ from Raman and FTIR experiments.